\documentclass[letterpaper, 11pt]{article}
\usepackage{graphicx}
\usepackage{amsmath}
\usepackage{amsfonts}
\usepackage{hyperref}
\usepackage{amsthm}
 \usepackage{amscd}
\usepackage{mathrsfs}
\usepackage{caption}
\usepackage{subcaption}
\usepackage{float}
\usepackage[top=1in,left=1in,right=1in,bottom=1in]{geometry}

\counterwithin{figure}{section}
\hypersetup{
	colorlinks = true,
	linkcolor  = red,
	citecolor  = blue,
	filecolor  = cyan,
	urlcolor   = magenta,}

\numberwithin{equation}{section}

\title{The $\mathbf N$-soliton solution formula for the Kaup--Kupershmidt equation\thanks{Dedicated to the memories of David J. Kaup and Boris A. Kupershmidt.}}
\author{Tuncay Aktosun\thanks{Corresponding author: aktosun@uta.edu}\\
Department of Mathematics\\
University of Texas at Arlington\\
Arlington, TX 76019-0408, USA\\
\\
Ramazan Ercan\\
Department of Mathematics\\
California State University San Marcos\\
San Marcos, CA 92096, USA\\
\\
Mehmet Unlu\\
Department of Mathematics\\
Recep Tayyip Erdogan University\\
53100 Rize, Turkey}

\date{}

\begin{document}

\maketitle

\begin{abstract}

An explicit formula is presented for the $\mathbf N$-soliton solution to the Kaup--Kupershmidt
equation, for any positive integer $\mathbf N,$ by using the input 
data set consisting of $\mathbf N$ distinct positive $p_j$-values 
and $\mathbf N$ arbitrarily chosen positive $c_j$-values. 
The formula is derived from the solution to the inverse 
scattering problem for a third-order linear operator in the reflectionless case 
with the $p_j$-values related to the bound-state poles of a transmission 
coefficient and the $c_j$-values related to the bound-state dependency constants.
It is indicated how the formula is constructed from the 
solution to a Riemann--Hilbert problem and 
from the solution to a Marchenko integral equation used in the formulation of the 
aforementioned inverse scattering problem.

\end{abstract}

{\bf {AMS Subject Classification (2020):}} 34A55 34M50 35C08

{{\bf Keywords:} Kaup--Kupershmidt equation,  $\mathbf N$-soliton solution formula, 
inverse scattering for the third-order equation, bound-state dependency constants, 
Riemann--Hilbert problem,
Marchenko integral equation}

\newpage

\section{Introduction}
\label{section1}

We consider
the Kaup--Kupershmidt equation
\begin{equation}
\label{1.1}
Q_t+Q_{xxxxx}+\displaystyle\frac{25}{2}\,Q_x\,Q_{xx}+5\,Q\,Q_{xxx}+5\,Q^2\,Q_x=0, \qquad x,\,t\in\mathbb R,
\end{equation}
where the quantity $Q$ is a real-valued function of
the spacial and temporal variables $x$ and $t,$ respectively, taking values on the real axis $\mathbb R,$ 
and the subscripts denote the appropriate partial derivatives. Like the KdV (Korteweg--de Vries) equation
\cite{KdV1895}, the Kaup--Kupershmidt equation \cite{K1980,K1984} is used to 
model propagation of shallow water waves, but 
by taking into consideration higher-order dispersive and nonlinear effects. 

Our primary goal  in this paper is to
 present the $\mathbf N$-soliton solution formula for \eqref{1.1}, for 
any positive integer $\mathbf N,$ where the solution 
contains $2\mathbf N$ arbitrary positive parameters, by solving the inverse scattering problem
in the reflectionless case for the corresponding third-order linear operator. We solve the relevant
inverse scattering problem by two methods. The first is achieved by formulating it as a Riemann--Hilbert problem 
and by recovering $Q$ in \eqref{1.1} from the solution to that Riemann--Hilbert problem. The second
is by establishing the corresponding Marchenko integral equation and by recovering $Q$ from the solution
to that integral equation. In either case, the $\mathbf N$-soliton solution to \eqref{1.1} is expressed in
an explicit form as the derivative of a logarithmic derivative of a particular function
of $x$ and $t$ containing $2\mathbf N$ positive parameters, where $\mathbf N$ of those parameters
are coming from the bound-state poles of a transmission coefficient and the remaining
$\mathbf N$ parameters coming from the bound-state dependency constants. Consequently, the explicit form
of our $\mathbf N$-soliton solution formula is expressed in 
the ``simplified version of Hirota's form," a term sometimes used
in the literature \cite{P2000}.

The significance and originality of our work are due to several reasons. 
An $\mathbf N$-soliton solution formula for the Kaup--Kupershmidt 
equation for an arbitrary positive integer $\mathbf N$
has not been available.
In the relevant paper by Parker \cite{P2000}, it is stated that ``exact soliton solutions of the Kaup--Kupershmidt
equation are derived using a simplified approach based in part on Hirota’s bilinear transform. These solutions have not
previously been obtained explicitly by analytic methods. It is found that, at each order, 
the $\mathbf N$-soliton solution is characterized
by an additional parameter which accounts for the `anomalous' nature of these solutions. 
The first four solitons are constructed
explicitly, and in a novel departure from Hirota’s method, an iterative procedure is 
presented for obtaining the general $\mathbf N$-soliton
solution.'' 

Parker \cite{P2000} also indicated that ``Of course, other standard analytic techniques 
for obtaining solutions of nonlinear evolution equations (NEEs) are
available \cite{AS1981}; but, for the Kaup--Kupershmidt equation, some have proved 
unsuccessful while others have not
been attempted. Using inverse scattering theory, Kaup \cite{K1980} was able to 
derive the solitary wave of the KK equation,
but no other soliton solutions; in the light of our results, this is perhaps understandable. 
Nevertheless, it seems to us
that obtaining the multisoliton solutions of the KK equation via IST, or one of its variants, remains a worthwhile
goal. In this connection, it is well to note that Hirota \cite{H1989} has 
solved the related Sawada--Kotera equation using
the inverse scattering scheme proposed by Zakharov and Shabat \cite{ZS1974}, 
but in doing so was forced to abandon the
pivotal Gel’fand--Levitan--Marchenko equation of IST \cite{AS1981}.''  

We remark that, in the quotation by Parker we have included above, Hirota's method refers to the ad-hoc method
 \cite{H2004} based on the use of Hirota's bilinear form, KK is the abbreviation for Kaup--Kupershmidt, 
 IST refers to the
inverse scattering transform \cite{GGKM1967}, 
the Sawada--Kotera equation \cite{SK1974} is the integrable evolution equation given by
\begin{equation}
\label{1.2}
Q_t+Q_{xxxxx}+5\,Q_x\,Q_{xx}+5\,Q\,Q_{xxx}+5\,Q^2\,Q_x=0, \qquad x,\,t\in\mathbb R,
\end{equation}
the inverse scattering scheme by Zakharov and Shabat refers to
the dressing method \cite{Z1990} of Shabat and Zakharov, and
the Gel’fand--Levitan--Marchenko equation refers to the Marchenko integral equation used
to solve various inverse scattering problems associated with the linear
second-order differential  operator \cite{CS1989,DT1979,F1967,L1987,M2011} 
and with some other first-order linear systems such as the AKNS system \cite{AKNS1974}.

Even though Parker outlined an iterative method \cite{P2000} to obtain 
soliton solutions to the Kaup--Kupershmidt equation
\eqref{1.1}, its implementation becomes extremely complicated 
and not practical beyond $\mathbf N=4.$ This is due to the fact that
the aforementioned iterative procedure uses not only 
a certain set of double-indexed coefficients $A_{jl}$ that are originally used \cite{CDG1976}
in the construction of the $\mathbf N$-soliton solution formula for the Sawada--Kotera equation, but 
it also requires the construction of an additional set of multi-indexed coefficients. The construction of those
additional multi-indexed coefficients \cite{P2000} becomes extremely complicated and not
practical as the value of $\mathbf N$ increases. Parker refers to soliton solutions to the
Kaup--Kupershmidt equation as ``anomalous soliton solutions'' and attributes 
the anomaly to the existence of those additional multi-indexed coefficients. 
In contrast, Parker refers to soliton solutions
to other evolution equations such as the KdV equation and
the Sawada--Kotera equation as ``regular soliton solutions.''

Our paper shows that, in order to express 
the $\mathbf N$-soliton solution formula for the Kaup--Kupershmidt equation
in a form resembling a ``simplified version of Hirota's form," one does not need to construct
an additional set of multi-indexed coefficients Parker \cite{P2000} aimed to construct. Instead
of using the original set of double-indexed coefficients $A_{jl},$ if one uses two distinct
sets of double-indexed coefficients we introduce
as $A^+_{jl}$ and $A^-_{jl},$ then no additional set of multi-indexed coefficients 
is needed. Hence,
there is no need for Parker's additional coefficients in order to express soliton solutions
in a simplified Hirota's form. Consequently, there is no need to use Parker's iterative procedure to construct
those additional coefficients. In summary, the complicated and
lengthy iterative procedure of Parker can completely be 
avoided in constructing soliton solutions to the Kaup--Kupershmidt equation \eqref{1.1}.

In \cite{MV2000}, Musette and Verhoeven mention that 
``As opposed to the Sawada–Kotera equation which is its dual equation, the construction of the
$\mathbf N$-soliton solutions of KK is not an easy task in using a perturbation scheme or the Hirota bilinear formalism. 
From the B\"acklund transformation obtained by singularity analysis considerations, we here establish the permutability theorem for the
KK equation. This allows us to explain the `anomalous' structure of the 
$\mathbf N$-soliton solution, empirically obtained by Parker \cite{P2000}."
It is indicated in \cite{MV2000} that 
``Then applying to KK a similar procedure, we obtain from its auto-BT the nonlinear superposition formula which
turns out to be a differential equation of second order and second degree of the Appell type which is linearizable.
Its solution leads to the permutability theorem for KK and to the explicit expressions (63) and (73) of the $2$- and
$3$-soliton tau-functions which in terms of the vacuum solutions of the Lax pair are determinants of the Gram type."
The explicit construction of that determinant is provided in (60) for
$\mathbf N=1$ and in (66) for $\mathbf N=2$ and the construction for
$\mathbf N=3$ is described in (67)--(73) in \cite {MV2000}.
It is mentioned in \cite{MV2000} that ``Now, expressions (63) and (73) suggest that the $\mathbf N$-parameter tau-function may be written as" in (75) of \cite{MV2000}, 
which is an $\mathbf N\times \mathbf N$ determinantal expression where
the diagonal terms and the off-diagonal terms need to be constructed from an appropriate input data set
such as $\{p_j,c_j\}_{j=1}^{\mathbf N}$ that we use in our paper.
The explicit closed-form evaluation of that $\mathbf N\times \mathbf N$ determinant is not
as simple or straightforward compared to our explicit determination of
$\tau$ appearing in our formula \eqref{2.16}, which is simply
and easily displayed as an explicit evaluation of a $2\mathbf N\times 2\mathbf N$ determinant for
an arbitrary positive integer $\mathbf N.$
As we have already indicated, unless one uses the double-indexed coefficients
 $A^+_{jl}$ and $A^-_{jl}$ we introduce, the evaluation of the
$\mathbf N$-soliton solution
to the Kaup--Kupershmidt equation
 with 
a determinantal expression does not yield
a simple closed-form expression that can be explicitly and easily written for an
arbitrary positive integer $\mathbf N.$

As we show in our paper, we can express
the $\mathbf N$-soliton solution formula, for an arbitrary positive integer $\mathbf N,$ 
for the Kaup--Kupershmidt equation
by using only the set of double-indexed coefficients $A^+_{jl}$ and $A^-_{jl}$
without needing a complicated and impractical iterative procedure.
We mention that the $1$-soliton solution formula for the Kaup--Kupershmidt equation
was first obtained by Kaup \cite{K1980}, the $2$-soliton and $3$-soliton solution formulas
were provided by Hereman and Nuseir \cite{HN1997}, and the $4$-soliton
solution formula was given by
Parker \cite{P2000}.
In the survey paper \cite{HG2024}, a Mathematica code is used by Hereman and Goktas to obtain
the $\mathbf N$-soliton solutions to 
 the Kaup--Kupershmidt equation for $\mathbf N=1,2,3$
utilizing the simplified Hirota's form. However, the ``growing complexity" beyond $\mathbf N=3$ is indicated
 in \cite{HG2024}
by mentioning that 
``The computations are very lengthy, time consuming, and currently at the limit of what Mathematica
can handle."
Our $\mathbf N$-soliton solution formula in Hirota's simplified form 
readily yields those aforementioned solutions
not only when $\mathbf N$ is chosen as $1,$ $2,$ $3,$ and $4,$ respectively, but
for any arbitrary positive integer $\mathbf N.$
For $1\le \mathbf N \le 10,$ we have prepared
Mathematica notebooks to plot and animate soliton solutions to the Kaup--Kupershmidt 
equation where the user can provide the input
$\{p_j,c_j\}_{j=1}^{\mathbf N}.$ Those Mathematica notebooks are available 
upon request from the first author. In Figures~\ref{figure6.1} and
\ref{figure6.2} in our paper, we present some snapshots for
the $7$-soliton solution and the $10$-soliton solution, respectively,
to the Kaup--Kupershmidt equation.

It is mentioned in the paper \cite{W2025} by Wang that the Kaup--Kupershmidt equation ``is notably 
challenging to analyze through conventional methods, in contrast to other integrable
systems. Unlike first and second-order spectral problems commonly 
solved in AKNS systems \cite{AKNS1974,ZS1972}, the
third-order spectral problem associated with (KK) is much more complex and its analysis remains largely
unresolved. Third-order spectral problems, such as those arising in 
the context of the Boussinesq equation
and the Caudrey--Dodd--Gibbon--Sawada--Kotera (CDGSK) equation are central 
topics in mathematical physics." Wang \cite{W2025} added that ``Kaup \cite{K1980} 
made partial progress on the inverse scattering problem associated with (KK), deriving
``anomalous" soliton solutions for the KK equation. However, the general $\mathbf N$-soliton solutions of (KK)
have not been fully determined."
We remark that the Caudrey--Dodd--Gibbon--Sawada--Kotera equation \cite{CDG1976} mentioned
in Wang's paper is another name used for the Sawada--Kotera equation \cite{SK1974}.

Another reason for the significance and originality of our work is due to the 
fact that we obtain the $\mathbf N$-soliton solution formula 
for the Kaup--Kupershmidt equation by solving the inverse scattering problem 
for the associated third-order linear differential operator, 
in which $Q$ appears as a coefficient. We solve the corresponding inverse scattering 
problem in the reflectionless case by specifying our input data set consisting 
of $\mathbf N$ bound-state poles of a transmission coefficient and $\mathbf N$ 
time-evolved bound state dependency constants. 
The solution method we use to construct soliton solutions to the Kaup--Kupershmidt 
equation is exactly the same as the method we used \cite{ACTU2025,ACTU2026}
to construct soliton solutions to the Sawada--Kotera equation. In fact, the same 
transmission coefficient and hence the same set of $\mathbf N$ bound-state
poles are used as input to construct each of the $\mathbf N$-soliton solution 
to the Kaup--Kupershmidt  equation and to the Sawada--Kotera
equation. The only difference is the set of bound-state dependency constants we use
to construct those soliton solutions. We elaborate on this issue later in our paper.
This shows that it is unnecessary to classify soliton solutions
as ``regular'' and ``anomalous.'' 
Since we use the same method to obtain soliton solutions to the
Kaup--Kupershmidt equation and to the Sawada--Kotera equation, we are also 
able to explain the cause in the ``difficulty'' of obtaining
 soliton solutions to the Kaup--Kupershmidt  equation compared to
soliton solutions to the Sawada--Kotera equation if an ad-hoc method is used
to construct those solutions. A key determinantal quantity in the Sawada--Kotera case becomes a perfect
square and this does not happen
in the Kaup--Kupershmidt case, and hence the formula for 
the $\mathbf N$-soliton solution for the Sawada--Kotera equation 
becomes much simpler.

Yet another reason for the significance and originality of our work is that we 
obtain the $\mathbf N$-soliton solution to the Kaup--Kupershmidt
 equation by establishing the Marchenko integral equation that Kaup 
relentlessly but unsuccessfully pursued \cite{K2002}.
In fact, the desire to obtain the analog of the Marchenko integral equation 
yielding solutions to the Sawada--Kotera equation and the Kaup--Kupershmidt  equation 
was expressed by Kaup in his 2002 paper \cite{K2002}. The last section of Kaup's paper \cite{K2002} has the title ``Unsolved Problems,''
where Kaup stated that ``Now we will discuss
what the hard problems are. They are hard because they have not been solved. This is not
a complete list, but is the start of a list of problems in need of solutions. Although we have a
solution for the IST of the cubic eigenvalue problem, there still is no equivalent of the GLM
equations for this system. What we do have is the solution of the Riemann–Hilbert problem
for the eigenfunctions. What is next needed is a representation of the eigenfunctions in terms
of some transformation kernels (like the $K(x, y; t)$ for the KdV). Once these are known, then
the equivalent GLM equations will follow.''
The GLM equation Kaup refers to is precisely the Marchenko integral equation  we present in our paper.
We recall that Parker \cite{P2000} refers to ``the pivotal Gel’fand--Levitan--Marchenko equation of IST''
and 
states that ``it seems to us that obtaining multi soliton solutions to the 
Kaup--Kupershmidt equation via IST, or one of its variance, remains a worthwhile goal.''
The absence of ``the GLM equation'' mentioned by Kaup
 \cite{K2002} and the lack of ``the pivotal Gel'fand--Levitan--Marchenko equation of IST'' 
mentioned by Parker \cite{P2000} are now remedied by the presentation of the Marchenko integral equation in our paper and 
 our use of its solution to construct the $\mathbf N$-soliton solution to the Kaup--Kupershmidt equation.
 In fact, the Marchenko equation we use to obtain soliton solutions to the
 Kaup--Kupershmidt equation is the same as the Marchenko equation \cite{AETU2026} we have used
 to obtain soliton solutions to the Sawada--Kotera equation. The only difference in 
these two cases is the input data differing
 from each other in the set of bound-state dependency constants.
 
Our paper is organized as follows. In Section~\ref{section2}, we present the $\mathbf N$-soliton 
solution formula \eqref{2.16} for the Kaup--Kupershmidt equation \eqref{1.1}
for any positive integer $\mathbf N.$ This is done by explicitly writing the function $\tau$ in \eqref{2.16}
as a polynomial of degree $2\mathbf N$ in the quantities $\chi_j$ in \eqref{2.7} with the coefficients expressed in terms of the quantities $A_{jl}^-$
and 
$A_{jl}^+$ in \eqref{2.21} and \eqref{2.22}, respectively. In Section~\ref{section3}, we describe 
how the key function $\tau$ is formed from the input 
data set $\{p_j,c_j\}_{j=1}^{\mathbf N}$ used to solve a 
particular inverse scattering problem for the third order equation \eqref{3.1} in the reflectionless case.
We indicate that the distinct positive $p_j$-values are related to the bound-state poles 
of the left transmission coefficient 
and the arbitrarily chosen positive $c_j$-values are related to the bound-state dependency constants.
In Section~\ref{section4}, we show how the key function $\tau$ is constructed from the solution to 
a Riemann--Hilbert problem associated with the aforementioned inverse scattering problem using the input data set $\{p_j,c_j\}_{j=1}^{\mathbf N}.$
In Section~\ref{section5}, we indicate how the key function $\tau$ is obtained from the solution to the Marchenko integral equation \eqref{5.1} associated with the related inverse scattering problem using the 
similar input data set $\{p_j,c_j\}_{j=1}^{\mathbf N}.$
Finally, in Section~\ref{section6}, we reiterate the significance of our main result in the paper, provide 
as two examples some snapshots for the $7$-soliton solution and
some snapshots for the $10$-soliton solution
to the Kaup--Kupershmidt equation constructed by our method, and mention the planned related research in the near future.

\section{The $\mathbf N$-soliton solution formula}
\label{section2}

In this section we present the $\mathbf N$-soliton solution formula for the 
Kaup--Kupershmidt equation \eqref{1.1}, for any positive integer
$\mathbf N,$ by explicitly expressing it in terms of $2\mathbf N$ arbitrary 
positive constants, $\mathbf N$ of which must be distinct. Even though we derive the formula by solving 
the inverse scattering problem associated with the third-order linear operator, in 
the present section we concentrate on the formula itself without 
showing the steps in the derivation methods. This is done intentionally so that our 
paper is accessible to a wider audience, especially to
those researchers in the field of integrable evolution equations without expertise 
in the area of inverse scattering theory for linear
differential operators.

Our $\mathbf N$-soliton solution formula for \eqref{1.1} in Hirota's simplified form is given as
\begin{equation}\label{2.1}
Q(x,t)=3\,\displaystyle\frac{\partial}{\partial x}\left[
\displaystyle\frac{1}{\tau(x,t)}\left(\displaystyle\frac{\partial \tau(x,t)}{\partial x}\right)\right],\qquad x,t\in\mathbb R,
\end{equation}
where $\tau(x,t)$ is a polynomial of degree $2\mathbf N$ in the $\mathbf N$ real-valued quantities
$\chi_1,$ $\chi_2,$ $\dots,$ $\chi_\mathbf N$ and each $\chi_j$ appears in that 
polynomial at most to the second power. We have
\begin{equation}\label{2.2}
\chi_j:=\exp\left[-\left(k_j+k_j^\ast\right)x-9\left(k_j^5+(k_j^\ast)^5\right)t\right], \qquad 1\le j\le \mathbf N,
\end{equation}
with the asterisk denoting complex conjugation and $k_j$ is the complex-valued constant given by
\begin{equation}\label{2.3}
k_j:=iz\eta_j, \qquad 1\le j\le \mathbf N.
\end{equation}
Here, $z$ is the special complex number $e^{2\pi i/3},$ which is equivalently expressed as
\begin{equation}\label{2.4}
z=-\displaystyle\frac{1}{2}+i\displaystyle\frac{\sqrt{3}}{2}.
\end{equation}
The quantities $\eta_j$ are distinct positive constants. Without loss of generality, we can order them as
\begin{equation}\label{2.5}
0<\eta_1<\eta_2<\dots<\eta_{\mathbf N}.
\end{equation}
Alternatively, instead of using $\eta_j,$ we can use the related positive constant $p_j$ defined as
\begin{equation}\label{2.6}
p_j:=\displaystyle\frac{\eta_j}{\sqrt{3}}, \qquad 1\le j\le \mathbf N.
\end{equation}
Using \eqref{2.3}--\eqref{2.6} in \eqref{2.2}, we see that we can write $\chi_j$ in a simple form as
\begin{equation}\label{2.7}
\chi_j=e^{p_j x-p_j^5 t}, \qquad 1\le j\le \mathbf N.
\end{equation}

We introduce the variables $y_j$ and $y_j^\ast$ as
\begin{equation}\label{2.8}
y_j:=c_j \chi_j, \qquad 1\le j\le \mathbf N,
\end{equation}
\begin{equation}\label{2.9}
y_j^\ast:=c_j^\ast \chi_j, \qquad 1\le j\le \mathbf N,
\end{equation}
where each $c_j$ is assumed to be an arbitrary complex constant so that 
we can make a distinction between $y_j$ and its complex conjugate
$y_j^\ast.$ In the $\mathbf N$-soliton solution formula we eventually 
choose each $c_j$ as an arbitrary positive constant so that $y_j$ and
$y_j^\ast$ eventually become equal to each other. However, in order to 
present our $\mathbf N$-soliton solution formula in the clearest
possible way, we initially need to make a distinction between $y_j$ and $y_j^\ast$ 
and hence we initially assume that each $c_j$ is 
an arbitrary complex parameter.

We introduce the set $\mathbf Y$ with $2\mathbf N$ elements as
\begin{equation}\label{2.10}
\mathbf Y:=\{y_1,y_1^\ast,\dots,y_{\mathbf N},y_{\mathbf N}^\ast\}.
\end{equation}
For $2\le l\le 2\mathbf N,$ we define the set $\mathbf Y_l$ consisting of $l$-tuple 
Cartesian product of the elements from $\mathbf Y$
in such a way that we exclude self-matching and ignore order. The exclusion of 
self-matching means that, in the set $\mathbf Y_l,$ each
element from $\mathbf Y$ can appear at most to the first power. By ignoring order, 
for example, we consider the pair $y_j\,y_l$ and
$y_l\,y_j$ to be the same and similarly we consider $y_j\,y_l^\ast$ and $y_l^\ast\,y_j$ 
to be the same. Hence, we include $y_j\,y_l$ but
exclude $y_l\,y_j$ when $j<l,$ include $y_j\,y_j^\ast$ but exclude $y_j^\ast\,y_j,$ 
include $y_j^\ast\,y_l^\ast$ but exclude
$y_l^\ast\,y_j^\ast$ when $j<l.$ Thus, each set $\mathbf Y_l$ for $2\le l\le 2\mathbf N$ 
contains precisely $\binom{2\mathbf N}{l}$ elements, where we use
$\binom{m}{n}$ to denote the binomial coefficient defined as
\begin{equation*}
\displaystyle\binom{m}{n}:=\displaystyle\frac{m!}{n!(m-n)!}, \qquad 1\le n\le m\le 2\mathbf N.
\end{equation*}
For example, we have
\begin{equation}\label{2.12}
\mathbf Y_2:=\{y_1 y_1^\ast,y_1 y_2,y_1 y_2^\ast,\dots,y_{\mathbf N-1}^\ast 
y_{\mathbf N}^\ast, y_{\mathbf N} y_{\mathbf N}^\ast\},
\end{equation}
\begin{equation*}
\mathbf Y_3:=\{y_1 y_1^\ast y_2,y_1 y_1^\ast y_2^\ast,\dots,y_{\mathbf N-1}^\ast 
y_{\mathbf N} y_{\mathbf N}^\ast\},
\end{equation*}
\begin{equation*}
\mathbf Y_4:=\{y_1 y_1^\ast y_2 y_2^\ast,y_1 y_1^\ast y_2 y_3,\dots,y_{\mathbf N-1} 
y_{\mathbf N-1}^\ast y_{\mathbf N} y_{\mathbf N}^\ast\},
\end{equation*}
\begin{equation}\label{2.15}
\mathbf Y_{2\mathbf N}:=\{y_1 y_1^\ast y_2 y_2^\ast \cdots y_{\mathbf N} y_{\mathbf N}^\ast\}.
\end{equation}
We see from \eqref{2.15} that $\mathbf Y_{2\mathbf N}$ contains a single element consisting of the product
of all elements in $\mathbf Y.$
For convenience, we define $\mathbf Y_1$ by letting $\mathbf Y_1:=\mathbf Y.$

Let us describe the quantity $\tau$ appearing in \eqref{2.1} by suppressing its $t$-dependence
in our notation. In the solution to the
relevant inverse scattering problem
for the third-order linear differential equation, 
the time variable $t$ appears not as an independent variable but as a parameter. Because of this,
we replace the partial $x$-derivative in \eqref{2.1} with the
ordinary $x$-derivative, and we write \eqref{2.1} as
\begin{equation}\label{2.16}
Q(x)=3\left(\displaystyle\frac{\tau'}{\tau}\right)',
\end{equation}
where the prime denotes the $x$-derivative. 
Our goal is to provide an explicit expression for $Q$ in terms of
$x,$ $t,$ and the parameters in the set
$\{p_j,c_j\}_{j=1}^{\mathbf N}$ so that
\eqref{2.16} constitutes the $\mathbf N$-solution formula for the
Kaup--Kupershmidt equation \eqref{1.1}. 
From \eqref{2.16} we see that
$Q$ is uniquely determined if $\tau$ is specified. Thus, for our goal, it is sufficient that 
we express $\tau$ as an explicit function 
$x,$ $t,$ and the parameters in the set
$\{p_j,c_j\}_{j=1}^{\mathbf N}.$

We remark that the quantity $\tau$ appearing in \eqref{2.16} is the counterpart 
of the auxiliary function
introduced by Hirota \cite{H1971} to express soliton solutions to the 
KdV (Korteweg-de Vries) equation as a derivative of the logarithmic
derivative of that auxiliary function. We refer to the form of such a formula 
as the simplified version of Hirota's form. We mention that
Hirota's auxiliary function, in the name of the $\tau$-function, was 
generalized by Date, Jimbo, Kashiwara, and Miwa to analyze soliton
solutions to the KP (Kadomtsev--Petviashvili) hierarchy \cite{DJKM1981} 
and to the BKP (B-type Kadomtsev--Petviashvili) hierarchy
\cite{DJKM1982}.

We write the quantity $\tau$ appearing in \eqref{2.16} as 
the summation of $2\mathbf N+1$ terms as
\begin{equation}\label{2.17}
\tau=1+\tau_1+\tau_2+\dots+\tau_{2\mathbf N},
\end{equation}
where we have defined $\tau_1$ as the sum of the elements 
in the set $\mathbf Y$ in \eqref{2.10}, i.e. we let
\begin{equation}\label{2.18}
\tau_1:=y_1+y_1^\ast+y_2+y_2^\ast+\dots+y_{\mathbf N}+y_{\mathbf N}^\ast.
\end{equation}
The term $\tau_2$ in \eqref{2.17} is obtained by taking the 
appropriate linear combination of the elements in the set $\mathbf Y_2$ in
\eqref{2.12}.
The coefficients contain $1/4$ and the double-indexed 
constants $A_{jl}^-$ and $A_{jl}^+,$ where we have defined 
\begin{equation}\label{2.19}
A_{jl}^-:=\displaystyle\frac{(k_j-k_l)^3}{k_j^3-k_l^3},\qquad 1\le j<l\le \mathbf N,
\end{equation}
\begin{equation}\label{2.20}
A_{jl}^+:=\displaystyle\frac{k_j^3+k_l^3}{(k_j+k_l)^3}, \qquad 1\le j<l\le \mathbf N.
\end{equation}
Using \eqref{2.3} and \eqref{2.6} in \eqref{2.19} and \eqref{2.20}, we get
\begin{equation}\label{2.21}
A_{jl}^-:=\displaystyle\frac{(\eta_j-\eta_l)^3}{\eta_j^3-\eta_l^3}, 
\quad A_{jl}^-:=\displaystyle\frac{(p_j-p_l)^3}{p_j^3-p_l^3},\qquad 1\le j<l\le \mathbf N,
\end{equation}
\begin{equation}\label{2.22}
A_{jl}^+:=\displaystyle\frac{\eta_j^3+\eta_l^3}{(\eta_j+\eta_l)^3}, 
\quad A_{jl}^+:=\displaystyle\frac{p_j^3+p_l^3}{(p_j+p_l)^3},\qquad 1\le j<l\le \mathbf N.
\end{equation}
Since the parameters $\eta_j$ for $1\le j\le \mathbf N$ are distinct, 
and hence also the parameters $p_j$ for $1\le j\le \mathbf N$ are
distinct, it follows from \eqref{2.21} and \eqref{2.22} that each of 
the constants $A_{jl}^-$ and $A_{jl}^+$ is positive. Furthermore, those
constants satisfy the symmetry relations
\begin{equation*}
A_{jl}^-=A_{lj}^-, \quad A_{jl}^+=A_{lj}^+, \qquad 1\le j<l \le \mathbf N.
\end{equation*}
We define $\tau_2$ appearing in \eqref{2.17} as
\begin{equation}\label{2.24}
\tau_2:=\displaystyle\frac{1}{4}\,y_1 y_1^\ast+A_{12}^- \,y_1 y_2+A_{12}^+\, y_1 
y_2^\ast+A_{13}^- \,y_1 y_3+A_{13}^+\, y_1\, y_3^\ast+\dots
+\displaystyle\frac{1}{4}\,y_{\mathbf N}\,y_{\mathbf N}^\ast.
\end{equation}
Thus, the coefficient of the product $y_j y_j^\ast$ is chosen as $1/4$ for $1\le j\le\mathbf N.$ For
$1\le j<l\le \mathbf N,$
the coefficient of $y_j y_l$ is $A_{jl}^-,$ the coefficient of $y_j^\ast  y_l^\ast$ is $A_{jl}^-,$ 
the coefficient of $y_j y_l^\ast$ is $A_{jl}^+,$ and the coefficient of $y_j^\ast y_l$ is $A_{jl}^+.$

In a similar way, we define $\tau_3$ appearing in \eqref{2.17} by using 
all linear combination of the elements in the set $\mathbf Y_3$ with
the appropriate choice of the coefficients. We get
\begin{equation}\label{2.25}
\begin{split}
\tau_3=\displaystyle\frac{1}{4}\,A_{12}^- \,A_{12}^+ \,y_1 \,y_1^\ast y_2+&
\displaystyle\frac{1}{4}\,A_{12}^+\, A_{12}^- \,y_1 \,y_1^\ast y_2^\ast+\displaystyle\frac{1}{4}\,A_{13}^- \,A_{13}^+ \,
y_1 \,y_1^\ast y_3+\cdots
\\
&+ A_{(\mathbf N-1)\mathbf N}^+\, A_{(\mathbf N-1)\mathbf N}^- \,\displaystyle\frac{1}{4}\,
y_{\mathbf N-1}^\ast\, y_{\mathbf N} \,y_{\mathbf N}^\ast.
\end{split}
\end{equation}
To obtain the coefficient of the first term $y_1 y_1^\ast y_2$ on the right-hand side of \eqref{2.25}, we proceed as follows. We use $1/4$
because of the pair product $y_1 y_1^\ast,$ use $A_{12}^-$ due to the pair product $y_1 y_2,$ and use $A_{12}^+$ due to the factor
$y_1^\ast y_2.$ The coefficient $(1/4)A_{12}^+ A_{12}^-$ of the term $y_1 y_1^\ast y_2^\ast$ is obtained because the first factor $1/4$ is
due to the product $y_1 y_1^\ast,$ the second factor $A_{12}^+$ is due to the product $y_1 y_2^\ast,$ and the third factor $A_{12}^-$ is
due to the product $y_1^\ast y_2^\ast.$ We use the similar pattern to determine the coefficients in 
the remaining terms. 
In a similar manner, we construct all the terms $\tau_l$ in \eqref{2.17} for $2\le l\le 2\mathbf N.$ 
For example, $\tau_{2\mathbf N}$ consists of the single term given by
\begin{equation}\label{2.26}
\tau_{2\mathbf N}:=\left(\displaystyle\frac{1}{4}\, A_{12}^- A_{12}^+ A_{13}^- A_{13}^+
\cdots A_{(\mathbf N-1)\mathbf N}^+ \,A_{(\mathbf N-1)\mathbf N}^-\,
\displaystyle\frac{1}{4}\right)\left(y_1\, y_1^\ast\, y_2\, y_2^\ast \cdots y_{\mathbf N-1}^\ast \,y_{\mathbf N} \,y_{\mathbf N}^\ast\right).
\end{equation}

After expressing each of $\tau_l$ for $1\le l\le 2\mathbf N$ in terms of the elements in the set $\mathbf Y,$ we use \eqref{2.8} and
\eqref{2.9} and write each $\tau_l$ in terms of $c_j\chi_j$ by restricting the parameters $c_j$ to arbitrary positive values. Hence, for the
$1$-soliton we have the value of $\tau$ appearing in \eqref{2.17} as
\begin{equation*}
\tau=1+\tau_1+\tau_2,
\end{equation*}
which yields
\begin{equation}\label{2.28}
\tau=1+y_1+y_1^\ast+\displaystyle\frac{1}{4}y_1 y_1^\ast.
\end{equation}
Using \eqref{2.8} and \eqref{2.9} in \eqref{2.28}, we obtain
\begin{equation}\label{2.29}
\tau=1+2c_1\chi_1+\displaystyle\frac{1}{4}c_1^2 \chi_1^2,
\end{equation}
where $\chi_1$ is the quantity defined in \eqref{2.7} and $c_1$ is an arbitrary positive parameter. Thus, using \eqref{2.29} in \eqref{2.16},
we express the $1$-soliton solution $Q$ to the Kaup--Kupershmidt equation \eqref{1.1} by using the input data set $\{p_1,c_1\}$ with
$p_1$ and $c_1$ being arbitrary positive parameters.

For the $2$-soliton solution, from \eqref{2.17} we get
\begin{equation}\label{2.30}
\tau=1+\tau_1+\tau_2+\tau_3+\tau_4,
\end{equation}
where we have
\begin{equation}\label{2.31}
\tau_1=y_1+y_1^\ast+y_2+y_2^\ast,
\end{equation}
\begin{equation}\label{2.32}
\tau_2=\displaystyle\frac{1}{4}\,y_1 y_1^\ast+A_{12}^- \,y_1y_2+A_{12}^+\, y_1y_2^\ast+A_{12}^+\, y_1^\ast y_2+A_{12}^- \,y_1^\ast y_2^\ast
+\displaystyle\frac{1}{4}\,y_2 y_2^\ast,
\end{equation}
\begin{equation}\label{2.33}
\tau_3=\displaystyle\frac{1}{4} A_{12}^- A_{12}^+\, y_1 y_1^\ast y_2
+ \displaystyle\frac{1}{4} A_{12}^+ A_{12}^- \,y_1 y_1^\ast y_2^\ast
+  A_{12}^- A_{12}^+\, \displaystyle\frac{1}{4}\,y_1 y_2 y_2^\ast 
+ A_{12}^+ A_{12}^- \,\displaystyle\frac{1}{4} \,y_1^\ast y_2 y_2^\ast,
\end{equation}
\begin{equation}\label{2.34}
\tau_4=\displaystyle\frac{1}{4} \,A_{12}^- A_{12}^{+} A_{12}^+ A_{12}^-\, \displaystyle\frac{1}{4}\, y_1 y_1^\ast y_2 y_2^\ast.
\end{equation}
Using \eqref{2.8} and \eqref{2.9} in \eqref{2.31}--\eqref{2.34}, we get
\begin{equation}\label{2.35}
\tau_1=2\left(c_1\chi_1+c_2\chi_2\right),
\end{equation}
\begin{equation}\label{2.36}
\tau_2=\displaystyle\frac{1}{4} \left( c_1^2 \chi_1^2 + c_2^2 \chi_2^2 \right) + 2 \left( A_{12}^- + A_{12}^+ \right) c_1 c_2 \chi_1 \chi_2,
\end{equation}
\begin{equation}\label{2.37}
\tau_3=\displaystyle\frac{1}{2} \,A_{12}^- A_{12}^+ \left( c_1^2 c_2 \chi_1^2 \chi_2 + c_1 c_2^2 \chi_1 \chi_2^2 \right),
\end{equation}
\begin{equation}\label{2.38}
\tau_4= \displaystyle\frac{1}{4} \left( A_{12}^- A_{12}^+ \right)^2 c_1^2 c_2^2 \chi_1^2 \chi_2^2.
\end{equation}
Thus, using \eqref{2.35}--\eqref{2.38} in \eqref{2.30} we express $\tau$ as a polynomial 
of degree $4$ in $\chi_1$ and $\chi_2$ with
the coefficients uniquely determined by the input data set $\{p_1,p_2,c_1,c_2\}$ 
consisting of $4$ arbitrary positive parameters
with $p_1$ and $p_2$ being distinct. Then,
using the resulting expression for $\tau$ in \eqref{2.16}, we obtain the $2$-soliton 
solution $Q$ to the Kaup--Kupershmidt equation \eqref{1.1}. 

In a similar manner, by using the procedure outlined, with the help of 
\eqref{2.8}, \eqref{2.9}, \eqref{2.16}--\eqref{2.18}, \eqref{2.21}, \eqref{2.22}, and 
\eqref{2.24}--\eqref{2.26}, we obtain the $\mathbf N$-soliton solution $Q$ to 
the Kaup--Kupershmidt equation \eqref{1.1} in terms of the input data set 
$\{p_j,c_j\}_{j=1}^{\mathbf N}$ containing $2\mathbf N$ arbitrary positive 
parameters with the parameters $p_j$ being distinct.
The expression for $\tau,$ even though expressed explicitly in terms of the 
input data set $\{p_j,c_j\}_{j=1}^{\mathbf N},$ becomes
lengthier as the value of $\mathbf N$ increases. Each term $\tau_l$ in \eqref{2.17} 
consists of the sum of $\binom{2\mathbf N}{l}$ terms.
Thus, the quantity $\tau$ consists of the sum of $2^{2\mathbf N}$ terms. Due to 
the symmetry $j\leftrightarrow l$ in $\tau,$ it is possible to
group similar terms and simplify the expression in each $\tau_l$ for $2\le l \le 2\mathbf N.$ 
Nevertheless, the number of terms in $\tau,$
even in its simplified form, becomes too large to display explicitly as $\mathbf N$ increases. 

Let us make a contrast between the $\mathbf N$-soliton solution formula for 
the Kaup--Kupershmidt equation \eqref{1.1} and the
$\mathbf N$-soliton solution formula for the
Sawada--Kotera equation \eqref{1.2}.
We see that \eqref{1.1} and \eqref{1.2} differ from each other by the coefficients in 
the third terms on their left-hand sides. The
derivation of the $\mathbf N$-soliton solution formula for \eqref{1.2} is 
accomplished \cite{ACTU2026} in a similar manner we obtain the
$\mathbf N$-soliton solution formula for \eqref{1.1}. Thus, the expression 
\eqref{2.16} also yields the $\mathbf N$-soliton solution formula
for the Sawada--Kotera equation. However, the terms $\tau_l$ for $2\le l \le 2\mathbf N$ 
appearing in \eqref{2.17} are different for the
Kaup--Kupershmidt equation and for the Sawada--Kotera equation. In the latter case, 
the quantity $\tau$ appearing in \eqref{2.16} and
\eqref{2.17} becomes a perfect square, i.e. we have
\begin{equation}\label{2.39}
\tau=\Delta^2,
\end{equation}
where $\Delta$ has the form
\begin{equation}\label{2.40}
\Delta=1+\Delta_1+\Delta_2+\dots+\Delta_{\mathbf N}.
\end{equation}
Using \eqref{2.39} in \eqref{2.16}, we can write the $\mathbf N$-soliton solution 
formula for the Sawada--Kotera equation \eqref{1.2} as
\begin{equation}\label{2.41}
Q(x)=6\left(\displaystyle\frac{\Delta'}{\Delta}\right)',
\end{equation}
The construction of each $\Delta_l$ in \eqref{2.40} is similar to the construction of 
$\tau$ in \eqref{2.17}. However, each $\Delta_l$ is
constructed by using only the quantities $y_1,y_2,\dots,y_{\mathbf N}$ 
appearing in \eqref{2.8} and \eqref{2.10} without
needing the remaining quantities $y_1^\ast,y_2^\ast,\dots,y_{\mathbf N}^\ast.$ 
Thus, the quantity $\Delta$ in \eqref{2.40} consists of the sum of
$2^{\mathbf N}$ terms only. Furthermore, $\Delta$ is a polynomial in $y_1,$ $y_2,$
$\dots,$ $y_{\mathbf N},$ where each term appears at most to
the first power. The coefficients in the $l$-tuples in the expression for $\Delta$ 
are constructed with the help of the coefficients $A_{jl}$ that
are related to $A_{jl}^-$ and $A_{jl}^+$ appearing in \eqref{2.19}--\eqref{2.22}. In fact, we have
\begin{equation}\label{2.42}
A_{jl}:=A_{jl}^- A_{jl}^+.
\end{equation}
We refer to (5.44)--(5.46) of \cite{AETU2026} for the details. For example, 
for the $1$-soliton solution for the Sawada--Kotera
equation \eqref{1.2}, we have
\begin{equation}\label{2.43}
\Delta_1=y_1,
\end{equation}
which is expressed, after using \eqref{2.8}, as
\begin{equation}\label{2.44}
\Delta_1=c_1\chi_1.
\end{equation}
For the $2$-soliton solution of the Sawada--Kotera equation, we get
\begin{equation}\label{2.45}
\Delta_1=y_1+y_2, \quad \Delta_2=A_{12}y_1 y_2,
\end{equation}
which yields, after using \eqref{2.8},
\begin{equation}\label{2.46}
\Delta_1=c_1 \chi_1+c_2 \chi_2, \quad \Delta_2=A_{12}c_1 c_2 \chi_1 \chi_2.
\end{equation}

In most of the terms in the quantities $\tau_l$ in \eqref{2.17}, $A_{jl}^-$ and $A_{jl}^+$ 
appear together as the product $A_{jl}^- A_{jl}^+.$
For example, from \eqref{2.36}--\eqref{2.38} we see that $A_{12}^-$ and $A_{12}^+$ do not 
appear in the multiplied form $A_{12}^- A_{12}^+$ in \eqref{2.36}.
This explains the motivation for Parker who tried \cite{P2000} to express all the terms 
$\tau_l$ in \eqref{2.17} in terms of the constants
$A_{jl}$ appearing in \eqref{2.42}. However, he was unable to construct those 
coefficients that could not be expressed in terms of $A_{jl}.$

\section{The inverse scattering problem}
\label{section3}

In Section~\ref{section2} we have presented the $\mathbf N$-soliton solution formula \eqref{2.16}
for the Kaup--Kupershmidt equation \eqref{1.1} by using the input data set 
$\{p_j,c_j\}_{j=1}^{\mathbf N}$
consisting of $2\mathbf N$ arbitrary positive parameters with the distinct $p_j$-values.
In this paper we show that the formula \eqref{2.16} actually comes from the solution to a relevant inverse scattering 
problem in the reflectionless case, where the parameters $p_j$ are related to the bound-state poles of a transmission coefficient
and the parameters $c_j$ are pertinent to the bound-state dependency constants.
In this section we describe the direct and inverse scattering problems associated with
the Kaup--Kupershmidt equation. Later, in Sections~\ref{section4} and \ref{section5}, we present
the solution to the relevant inverse problem
and show how \eqref{2.16} arises from the solutions of the corresponding Riemann--Hilbert problem and the
corresponding Marchenko integral equation, respectively.

The method we use to determine the $\mathbf N$-soliton solution formula, for an arbitrary positive integer $\mathbf N,$ is based on solving the inverse
scattering problem for the third-order equation
\begin{equation}
\label{3.1}
\psi'''+Q\,\psi'+P\,\psi=k^3\,\psi, \qquad x\in\mathbb R,
\end{equation}
where $x$ is the independent variable, the prime denotes the $x$-derivative, $k^3$ is the spectral parameter, $\psi$ is the wavefunction,
and the coefficients $Q$ and $P$ are the potentials that are functions of $x$ and contain $t$ as a parameter. For simplicity, we suppress
the $t$-dependence in the potentials and other quantities associated with the third-order equation \eqref{3.1}.

The direct scattering problem for \eqref{3.1} consists of the determination of the scattering coefficients and the bound-state information
when the potentials $Q$ and $P$ are specified in a special class of functions. On the other hand, the inverse scattering problem for \eqref{3.1}
comprises the determination of the potentials $Q$ and $P$ from an appropriate data set consisting of the scattering coefficients and the
bound-state information.

There are three special cases of \eqref{3.1}, where we have
\begin{equation}
\label{3.2}
P\equiv 0, \quad P=Q', \quad P=\displaystyle\frac{1}{2}\,Q',
\end{equation}
respectively. Viewing \eqref{3.1} in the operator form as $L\psi=k^3 \psi,$ we see that those three special cases correspond to the linear operators
$L_1,$ $L_2,$ $L_3,$ respectively, given by
\begin{equation}
\label{3.3}
L_1:=D^3+QD, \quad L_2:=D^3+QD+Q', \quad L_3:=D^3+QD+\displaystyle\frac{1}{2}\,Q',
\end{equation}
where we have defined $D:=d/dx$ and $D^3:=d^3/dx^3.$ Let us assume that the time dependence of the potentials $Q$ and $P$ in \eqref{3.1},
and hence the time dependence of the operators $L_1,$ $L_2,$ $L_3,$ in \eqref{3.3}, is governed by the linear operators $A_1,$ $A_2,$ $A_3,$ respectively, where we have let
 \begin{equation*}
 \begin{cases}
A_1:=9\,D^5+15\,Q D^3+15\,Q_x\,D^2+\big(10\,Q_{xx}+5\,Q^2\big)D,\\
\noalign{\medskip}
A_2:=9\,D^5+15\,Q D^3+\big(15\,P+15\,Q_x\big)D^2+\big(15\,P_x+10\,Q_{xx}+5\,Q^2\big)D
+\big(10\,P_{xx}+10\,Q \,P\big),\\
\noalign{\medskip}
A_3:=9\,D^5+15\,Q D^3+\displaystyle\frac{45}{2}\,Q_x\,D^2+\left(\displaystyle\frac{35}{2}\,Q_{xx}+5\,Q^2\right)D+
5\,Q_{xxx}+5\,Q\,Q_x.
\end{cases}      
\end{equation*}
Then, the three Lax pairs $(L_1,A_1),$ $(L_2,A_2),$ $(L_3,A_3),$ each yield the integrable evolution equations obtained by using the Lax
operator identity \cite{L1968}, i.e.
\begin{equation}
\label{3.5}
\dot L_j+L_j\,A_j-A_j\,L_j=0, \qquad j=1,2,3,
\end{equation}
where the overdot in $\dot L_j$ denotes the time derivative of the operator $L_j.$ Thus, we have $\dot L_1,$ $\dot L_2,$ $\dot L_3$ given,
respectively, as
 \begin{equation*}
\dot L_1=Q_t D, \quad \dot L_2=Q_t D+Q_{xt}, \quad \dot L_3=Q_t D+\displaystyle\frac{1}{2}Q_{xt},
\end{equation*}
where we recall that the subscripts $x$ and $t$ denote the appropriate partial derivatives.
Using the Lax pairs $(L_1,A_1)$ and $(L_2,A_2)$ in the operator equality \eqref{3.5}, we obtain the Sawada--Kotera equation \eqref{1.2}, and
by using the Lax pair $(L_3,A_3)$ in \eqref{3.5}, we obtain the Kaup--Kupershmidt equation \eqref{1.1}.
 
It may seem awkward that the Lax pairs $(L_1,A_1)$ and $(L_2,A_2)$ yield the same integrable evolution equation \eqref{1.2}. We refer the reader
to \cite{ACTU2025} to see why $(L_1,A_1)$ and $(L_2,A_2)$ correspond to the same Sawada--Kotera equation with the same time-evolved
potentials $Q$ but two different time-evolved potentials $P\equiv 0$ and $P=Q_x.$ If we exclusively work in the nonlinear domain, then we are
mainly concerned with the analysis of the initial-value problem for the Sawada--Kotera equation \eqref{1.2}. That initial-value problem involves
the determination of $Q$ for all $x\in\mathbb R$ and $t>0$ when we know $Q$ for all $x\in\mathbb R$ and at $t=0.$ Thus, unless we analyze the
scattering and inverse scattering problems for the linear operators $L_1$ and $L_2,$ respectively, we do not need to make a distinction between
the Lax pairs $(L_1,A_1)$ and $(L_2,A_2).$ On the other hand, the scattering and inverse scattering problems associated with those two Lax pairs
do not have identical solutions. If we deal with $(L_1,A_1),$ then the corresponding scattering data set is not necessarily the same as the
scattering data set associated with $(L_2,A_2).$ This is because in the former case we have $P\equiv 0$ and in the latter case we have $P=Q_x.$
We use SK1 to refer to the former case and use SK2 to refer to the latter case.

There is one special circumstance where the scattering data set for SK1 and the scattering data set for SK2 are closely related to each other.
That special circumstance occurs when the reflection coefficients are all zero so that the relevant scattering data set in each case is
determined by a left transmission coefficient, the bound-state poles of that transmission coefficient, and the dependency constants for the
bound states. It is known \cite{ACTU2025,ACTU2026} that in that special circumstance, i.e. in the reflectionless case, the two Lax pairs
$(L_1,A_1)$ and $(L_2,A_2)$ correspond to the same left transmission coefficient and hence the same set of bound states. The only difference
between the two Lax pairs is in the bound-state dependency constants. 

We can obtain the $\mathbf N$-soliton solution to the Kaup--Kupershmidt equation \eqref{1.1} in two different ways. One way is to choose the input
data set $\{p_j,c_j\}_{j=1}^{\mathbf N}$ for the relevant inverse scattering problem so that the bound-state constants $c_j$ are arbitrary positive
constants compatible with the third equality in \eqref{3.2}. The second way is to use the input data set $\{p_j,E_j\}_{j=1}^{\mathbf N},$ where each
positive bound-state constant $c_j$ is replaced with an arbitrary complex-valued dependency constant $E_j$ without initially imposing the restriction given in
the third equality in \eqref{3.2}. Next, we solve the relevant inverse scattering problem by using the input data set $\{p_j,E_j\}_{j=1}^{\mathbf N}.$ The
solution to the aforementioned inverse problem consists of the two time-evolved potentials $Q$ and $P.$ We then impose the restriction given in the
third equality in \eqref{3.2}. The resulting restriction also yields the restriction on the complex-valued dependency constants $E_j$ so that the
corresponding input data set is expressed as $\{p_j,c_j\}_{j=1}^{\mathbf N},$ where the positive bound-state constants $c_j$ are now compatible with
the third equality in \eqref{3.2}.

The advantage of proceeding as in the second way is the following. 
By using the input data set $\{p_j,E_j\}_{j=1}^{\mathbf N},$ we obtain the
$\mathbf N$-soliton solution to the integrable system given by 
\begin{equation}
\label{3.7}
\begin{cases}
Q_t+Q_{xxxxx}+5\,Q_x\,Q_{xx}+5\,Q\, Q_{xxx}+5\,Q^2\,Q_x
+15\,Q_{xx}\,P+15\,Q_x\, P_x-30\,P \,P_x=0,\\
\noalign{\medskip}
 \begin{aligned}
P_t+P_{xxxxx}+5\,Q\, P_{xxx}+15\,Q_x \,P_{xx}+20\,Q_{xx}\,P_x+&5\,Q^2\,P_x+10\,Q_{xxx}\,P-15\,P \,P_{xx}\\
&+10\,Q\, Q_x P-15\,(P_x)^2=0.
\end{aligned}        
\end{cases}
\end{equation}
One can directly verify that the three restrictions in \eqref{3.2} result in the reduction of the
system \eqref{3.7} to \eqref{1.2} in the SK1 case, to  \eqref{1.2} in the SK2 case, and
to \eqref{1.1}, respectively.
The $\mathbf N$-soliton solution to \eqref{3.7} consists of the complex-valued time-evolved potentials $Q$ and $P,$ where $P$ is not yet
related to $Q$ as in \eqref{3.2}. On the complex-valued $\mathbf N$-soliton solution
to \eqref{3.7}, by imposing the three restrictions in \eqref{3.2}, we
obtain the real-valued $\mathbf N$-soliton solution $Q$ to the Sawada--Kotera equation \eqref{1.2} in the SK1 case, the $\mathbf N$-soliton solution
$Q$ to \eqref{1.2} in the SK2 case, and the $\mathbf N$-soliton solution $Q$ to the Kaup--Kupershmidt equation \eqref{1.1}, respectively. We remark
that this procedure has been used in \cite{AETU2026} to obtain the $\mathbf N$-soliton solution to the Sawada--Kotera equation with the input data set
$\{p_j,c_j\}_{j=1}^{\mathbf N}.$ In our current paper, we use this method to obtain the $\mathbf N$-soliton solution to the Kaup--Kupershmidt equation.

As indicated (1.2)--(1.5) of \cite{ACTU2025}, the integrable system \eqref{3.7} is associated
with the Lax pair $(L,A)$ given by
 \begin{equation}
 \label{3.8}
L:=D^3+QD+P,
\end{equation}
\begin{equation}
 \label{3.9}
A:=9\,D^5+15\,Q D^3+\big(15\,P+15\,Q_x\big)D^2+\big(15\,P_x+10\,Q_{xx}+5\,Q^2\big)D
+\big(10\,P_{xx}+10\,Q \,P\big).
 \end{equation}
By using \eqref{3.8} and \eqref{3.9} in the Lax compatibility equation
\begin{equation*}
L_t+LA-AL=0,
\end{equation*}
we obtain the integrable system \eqref{3.7}.
 
From \eqref{2.3}--\eqref{2.6}, we observe that the parameters $p_j$ and $k_j$ are related to each other as
\begin{equation}
\label{3.11}
p_j=-\displaystyle\frac{iz^2}{\sqrt{3}}\,k_j, \qquad 1\le j\le \mathbf N.
 \end{equation}
Hence, the input data set $\{p_j,E_j\}_{j=1}^{\mathbf N}$ we intend to use to 
obtain the $\mathbf N$-soliton solution to the integrable system \eqref{3.7} is
equivalent to $\{k_j,E_j\}_{j=1}^{\mathbf N}.$ As described in Sections~3 and 4 of 
\cite{AETU2026}, in order to obtain the $\mathbf N$-soliton solution
to \eqref{3.7} with the input data set $\{k_j,E_j\}_{j=1}^{\mathbf N},$ one can use
the following steps:

\begin{enumerate}
    \item[\text{\rm{(a)}}] By using the $k_j$-values from the input data set, we form 
the left transmission coefficient $T_{\text{\rm{l}}}(k)$ for \eqref{3.1} in the reflectionless case.
As in (4.1) of \cite{AETU2026}, we get
\begin{equation*}
T_{\text{\rm{l}}}(k)=\displaystyle\frac{\Gamma(k)}{\Gamma(-k)}, \qquad k\in\mathbb C,
 \end{equation*}
where $\Gamma(k)$ is defined as
\begin{equation}
 \label{3.13}
\Gamma(k):=\displaystyle\prod_{j=1}^{\mathbf N} (k+k_j)(k+k_j^\ast), \qquad k\in\mathbb C.
 \end{equation}
We recall that the asterisk is used to denote complex conjugation.
The poles of $T_{\text{\rm{l}}}(k)$ occur at $k=k_j$ and $k=k_j^\ast$ for $1\le j\le\mathbf N,$
and they correspond to the bound-state poles because \eqref{3.1} has \cite{ACTU2025} a square-integrable solution
at $k=k_j$ and $k=k_j^\ast$ for $1\le j\le\mathbf N.$

\item[\text{\rm{(b)}}] Next, by using the $k_j$-values, we form the real-valued quantities 
$\chi_j$ defined in \eqref{2.2}. For the definition of the bound-state
dependency constants $E_j,$ we refer the reader to (4.5)--(4.7) of \cite{AETU2026}. 
We remark that $E_j$ is the analog of the bound-state dependency
constant arising in the direct and inverse scattering problems 
\cite{CS1989,DT1979,F1967} for the
Schr\"odinger equation on the full line. As indicated in (1.3) of \cite{AK2001}, a dependency
constant corresponds to the ratio of the left Jost solution to the right Jost solution
at a bound state of the corresponding Schr\"odinger operator.
As seen from (4.5)--(4.7) of \cite{AETU2026},
the constant $E_j$ corresponds to the ratio of the left Jost solution to the right Jost solution
at the bound state at $k=k_j$ for the third-order equation \eqref{3.1}.
Since the left Jost solution and the right Jost solution become linearly dependent
at a bound-state $k$-value, it is appropriate to refer to their ratio
as a bound-state dependency constant. The time evolution of the
bound-state dependency constant $E_j$ is indicated in the first equality
in (4.7) of \cite{AETU2026}. For the detailed information on
the bound states of \eqref{3.1} and
the determination of the
time evolution of
the corresponding bound-state dependency constants, we refer the reader
to Section~3 of \cite{ACTU2025}.

\item[\text{\rm{(c)}}] In the solution of the inverse scattering problem
for \eqref{3.1}, it is convenient \cite{ACTU2025,AETU2026} to use the modified 
bound-state dependency constants $\gamma_j$ instead of the 
bound-state dependency constants $E_j,$ where $\gamma_j$ is related
to $E_j$ as 
\begin{equation}
 \label{3.14}
\gamma_j:=-E_j\displaystyle\frac{\Gamma(k_j)}{\Gamma(-zk_j)}, \qquad 1\le j\le \mathbf N.
 \end{equation}
We recall that $z$ appearing in \eqref{3.14} is the special complex number in \eqref{2.4} and $\Gamma(k)$ is the quantity defined in \eqref{3.13}.
From \eqref{3.14} it is clear that if $E_j$ is an arbitrary complex parameter, then $\gamma_j$ also becomes an
arbitrary complex parameter.

\item[\text{\rm{(d)}}] Having the input data set $\{k_j,\gamma_j\}_{j=1}^{\mathbf N},$ we then solve the 
inverse scattering problem for \eqref{3.1}, and
we obtain the potentials $Q$ and $P$ with their time dependence governed by the integrable system \eqref{3.7}.
\end{enumerate}

The procedure outlined above yields the $\mathbf N$-soliton solution to \eqref{3.7} consisting of 
the time-evolved potential pair $Q$ and $P$ explicitly
expressed in terms of the parameters $k_j$ and $\gamma_j$ appearing in the data set $\{k_j,\gamma_j\}_{j=1}^{\mathbf N}.$ In order to obtain the
$\mathbf N$-soliton solution to the Kaup--Kupershmidt equation \eqref{1.1}, we then impose the restriction given in the third equality in \eqref{3.2}. That
restriction also results in the appropriate restriction on the complex-valued modified bound-state dependency constants $\gamma_j,$ which in turn yields
the arbitrary positive bound-state constants $c_j$ compatible with \eqref{1.1}. The resulting time-evolved potential $Q$ expressed in terms of the
parameters $k_j$ and $c_j$ in the input data set $\{k_j,c_j\}_{j=1}^{\mathbf N},$ or equivalently in 
terms of the parameters in $\{p_j,c_j\}_{j=1}^{\mathbf N},$ yields the explicit formula
\eqref{2.16} for the $\mathbf N$-soliton solution to the Kaup--Kupershmidt equation \eqref{1.1}. 

The complex-valued $\mathbf N$-soliton solution consisting of the time-evolved 
potential pair $Q$ and $P$ for the integrable system \eqref{3.7} is already explicitly constructed by using
the input data set $\{k_j,\gamma_j\}_{j=1}^{\mathbf N}.$ We refer the reader to Sections ~4 and 5 of \cite{AETU2026} for the aforementioned
construction by solving the corresponding inverse problem for \eqref{3.1} via the formulation of a related 
Riemann--Hilbert problem. The construction of the
$\mathbf N$-soliton solution to \eqref{3.7} by solving the corresponding inverse scattering problem 
for \eqref{3.1} via the formulation of a related Marchenko integral
equation can be found in Section~6 of \cite{AETU2026}.
In the next two sections, we describe the derivation of the fundamental formula \eqref{2.16} via the Riemann--Hilbert problem approach and the
Marchenko integral equation approach, respectively, tailored for the Kaup--Kupershmidt equation \eqref{1.1}.

\section{The Riemann--Hilbert problem}
\label{section4}

In this section we present the derivation of the $\mathbf N$-soliton solution formula \eqref{2.16} by solving the inverse scattering problem for \eqref{3.1}
associated with the Kaup--Kupershmidt equation \eqref{1.1}. As indicated in Section~\ref{section3}, we first obtain the $\mathbf N$-soliton solution formula
for the integrable system \eqref{3.7} via the Riemann--Hilbert problem formulation. That complex-valued $\mathbf N$-soliton solution to \eqref{3.7} is
obtained by using the input data set $\{k_j,\gamma_j\}_{j=1}^{\mathbf N},$ where the modified bound-state dependency constants $\gamma_j$ are
arbitrary complex-valued parameters. We then show how the condition given in the third equality in \eqref{3.2} restricts each $\gamma_j$ so that the input
data set can be presented as the data set $\{k_j,c_j\}_{j=1}^{\mathbf N},$ where the arbitrary 
positive bound-state constants $c_j$ are related to $\gamma_j$
compatible with the third equality in \eqref{3.2}. Consequently, the resulting $\mathbf N$-soliton 
solution formula for \eqref{3.7} yields the real-valued
$\mathbf N$-soliton solution formula \eqref{2.16} for \eqref{1.1}.

The formulation of the inverse scattering problem for \eqref{3.1} as a 
Riemann--Hilbert problem with the input data set $\{k_j,\gamma_j\}_{j=1}^{\mathbf N}$
is already available in Example~5.2
of \cite{ACTU2025} and in Sections~4 and 5 of \cite{AETU2026}. Instead of providing the formulation and the derivation of the solution of the aforementioned
Riemann--Hilbert problem, we refer the reader to \cite{ACTU2025,AETU2026}, and we start 
with the time-evolved potentials $Q$ and $P$ obtained from
the solution to that Riemann--Hilbert problem by using the input data set $\{k_j,\gamma_j\}_{j=1}^{\mathbf N}.$ 

We recall that the input data set $\{p_j,\gamma_j\}_{j=1}^{\mathbf N}$ is equivalent to the 
set $\{k_j,\gamma_j\}_{j=1}^{\mathbf N},$ where the distinct
positive $p_j$-values are related to the $k_j$-values as in \eqref{3.11}. We outline the recovery of $Q$ and $P$ in terms of the input data set
$\{k_j,\gamma_j\}_{j=1}^{\mathbf N}$ in the following step:

\begin{enumerate}
   \item[\text{\rm{(i)}}] The time-evolved potential Q is explicitly expressed in terms of the 
determinant of the $2\mathbf N\times 2\mathbf N$ matrix
$\mathbf M(x),$ where, as in (4.22) of \cite{AETU2026}, we have defined
\begin{equation}\label{4.1}
\mathbf M(x):=\begin{bmatrix}
m_{2\mathbf N-1}(k_1) & m_{2\mathbf N-2}(k_1) & \cdots & m_1(k_1) & m_0(k_1)\\
m_{2\mathbf N-1}(k_1^\ast) & m_{2\mathbf N-2}(k_1^\ast) & \cdots & m_1(k_1^\ast) & m_0(k_1^\ast)\\
\vdots & \vdots & \ddots & \vdots & \vdots \\
m_{2\mathbf N-1}(k_{\mathbf N}) & m_{2\mathbf N-2}(k_{\mathbf N}) & \cdots & m_1(k_{\mathbf N}) & m_0(k_{\mathbf N})\\
m_{2\mathbf N-1}(k_{\mathbf N}^\ast) & m_{2\mathbf N-2}(k_{\mathbf N}^\ast) & \cdots 
& m_1(k_{\mathbf N}^\ast) & m_0(k_{\mathbf N}^\ast)
\end{bmatrix}.
\end{equation}
We remark that the entries $m_l(k_j)$ and $m_l(k_j^\ast)$ in \eqref{4.1} are defined as in (4.24) of \cite{AETU2026}, i.e.
\begin{equation*}
\begin{cases}
m_l(k_j):=k_j^l+(zk_j)^l\,\gamma_j\,\chi_j,
\qquad 1\le j\le \mathbf N, \quad 0\le l\le 2\mathbf N,
\\
\noalign{\medskip}
m_l(k_j^\ast):=(k_j^\ast)^l+(z^2k_j^\ast)^l\,\gamma_j^\ast\,\chi_j,\qquad 1\le j\le \mathbf N, \quad 0\le l\le 2\mathbf N,
\end{cases}
\end{equation*}
where we recall that $z$ is the special complex-valued constant in \eqref{2.4} and $\chi_j$ is
the quantity defined in \eqref{2.2}.
Hence, it is clear that both $\mathbf M(x)$ and its determinant $\det[\mathbf M(x)]$ are explicitly determined 
by the input data set $\{k_j,\gamma_j\}_{j=1}^{\mathbf N}.$

\item[\text{\rm{(ii)}}] The time-evolved potential $Q$ is expressed in terms of $\det[\mathbf M(x)]$ and $\det[\mathbf M_2(x)],$ where $\mathbf M_2(x)$
is the $2\mathbf N\times 2\mathbf N$ matrix obtained by replacing the second column of the matrix $\mathbf M(x)$ given in \eqref{4.1} by the
$2\mathbf N$-column vector $\mathbf B(x)$ given by
\begin{equation}\label{4.3}
\mathbf B(x):=\begin{bmatrix}
m_{2\mathbf N}(k_1)\\
m_{2\mathbf N}(k_1^\ast)\\
\vdots\\
m_{2\mathbf N}(k_{\mathbf N})\\
m_{2\mathbf N}(k_{\mathbf N}^\ast)
\end{bmatrix}.
\end{equation}
We observe that each entry in $\mathbf B(x)$ is determined by the quantities $m_{2\mathbf N}(k_j)$ and $m_{2\mathbf N}(k_j^\ast)$ for
$1\le j\le \mathbf N,$ and hence $\det[\mathbf M_2(x)]$ itself is also explicitly determined by the input data set $\{k_j,\gamma_j\}_{j=1}^{\mathbf N}.$

\item[\text{\rm{(iii)}}] As indicated in (5.6) of \cite{AETU2026}, we have
 \begin{equation}\label{4.4}
Q(x)=3\,\left(\displaystyle\frac{(\det[\mathbf M(x)])'}{\det[\mathbf M(x)]}\right)',
\end{equation} 
where we recall that the prime denotes the $x$-derivative.

 \item[\text{\rm{(iv)}}] As indicated in (4.31) and (4.32) of \cite{AETU2026}, the potentials $Q$ and $P$ are given by
\begin{equation}\label{4.5}
Q(x)=-3\,A'_{2\mathbf N-1}(x),
\end{equation}
\begin{equation}\label{4.6}
P(x)=3\left[A_{2\mathbf N-1}(x)A'_{2\mathbf N-1}(x)-A''_{2\mathbf N-1}(x)-A'_{2\mathbf N-2}(x)\right],
\end{equation}
where we have
 \begin{equation}\label{4.7}
A_{2\mathbf N-1}(x)=-\Sigma_{\mathbf N}-\displaystyle\frac{(\det[\mathbf M(x)])'}{\det[\mathbf M(x)]},
\end{equation}
\begin{equation}\label{4.8}
A_{2\mathbf N-2}(x)=-\displaystyle\frac{\det[\mathbf M_2(x)]}{\det[\mathbf M(x)]},
\end{equation} 
with the constant $\Sigma_{\mathbf N}$ is defined as
\begin{equation}\label{4.9}
\Sigma_{\mathbf N}:=k_1+k_1^\ast+k_2+k_2^\ast+\dots+k_{\mathbf N}+k_{\mathbf N}^\ast.
\end{equation} 

\end{enumerate}
  
So far, we have determined the $\mathbf N$-soliton solution for \eqref{3.7} by expressing
the time-evolved potentials $Q$ and $P$ in terms of the input data set $\{k_j,\gamma_j\}_{j=1}^{\mathbf N}.$ Our goal now is
to apply the constraint $P=Q_x/2$ on the equalities \eqref{4.5} and \eqref{4.6} in order to determine the restriction on the modified
dependency constants $\gamma_j$ so that the resulting $Q$ in \eqref{4.4} yields the $\mathbf N$-soliton solution 
formula \eqref{2.16} for the Kaup--Kupershmidt
equation \eqref{1.1}. We remark that a similar procedure has been used in Section~5 of \cite{AETU2026} to 
determine the restrictions on $\gamma_j$ so that
$P=0$ in the SK1 case and $P=Q'$ in the SK2 are satisfied. Those two restrictions 
have yielded the solutions to the $\mathbf N$-soliton solution to the Sawada--Kotera
equation \eqref{1.2} in the SK1 and SK2 cases, respectively. 
We proceed as follows:

\begin{enumerate}
    
 \item[\text{\rm{(a)}}] We apply the restriction $P=Q'/2$ on \eqref{4.5} and \eqref{4.6}, and we obtain the constraint
 \begin{equation}\label{4.10}
3\left[A_{2\mathbf N-1}(x)\,A'_{2\mathbf N-1}(x)-A''_{2\mathbf N-1}(x)-A'_{2\mathbf N-2}(x)\right]=-\displaystyle\frac{3}{2}A''_{2\mathbf N-1}(x),
\end{equation} 
which simplifies to
 \begin{equation}\label{4.11}
A_{2\mathbf N-1}(x)\,A'_{2\mathbf N-1}(x)-\displaystyle\frac{1}{2}A''_{2\mathbf N-1}(x)-A'_{2\mathbf N-2}(x)=0.
\end{equation} 
The spacial asymptotics of $A_{2\mathbf N-1}(x),$ $A'_{2\mathbf N-1}(x),$ and $A_{2\mathbf N-2}(x)$ as $x\to +\infty$ are available in (4.33) of \cite{AETU2026},
and we have
\begin{equation}\label{4.12}
\displaystyle\lim_{x\to+\infty} A_{2\mathbf N-1}(x)=\Sigma_{\mathbf N},\quad
\displaystyle\lim_{x\to+\infty} A'_{2\mathbf N-1}(x)=0,\quad
\displaystyle\lim_{x\to+\infty} A_{2\mathbf N-2}(x)=\Sigma_{\mathbf N}^2-\Pi_{\mathbf N},
\end{equation}
with the constant $\Pi_{\mathbf N}$ defined as
\begin{equation}\label{4.13}
\begin{split}
\Pi_{\mathbf N}:=&k_1\left(k_1+k_1^\ast+\cdots+k_{\mathbf N}
+k_{\mathbf N}^\ast \right)+k_1^\ast\left(k_1^\ast+k_2+\cdots+k_{\mathbf N}+k_{\mathbf N}^\ast \right)\\
&+k_2\left(k_2+k_2^\ast+\cdots+k_{\mathbf N}+k_{\mathbf N}^\ast \right)+\cdots
+k_{\mathbf N}\left(k_{\mathbf N}+k_{\mathbf N}^\ast\right)+ k_{\mathbf N}^\ast\left(k_{\mathbf N}^\ast\right).
\end{split}
\end{equation}
Using \eqref{4.12}, we integrate \eqref{4.11} over the interval $(x,+\infty)$ and obtain the constraint equivalent to \eqref{4.10}. We get 
\begin{equation}\label{4.14}
\displaystyle\frac{1}{2}A_{2\mathbf N-1}(x)^2-\displaystyle\frac{1}{2}A'_{2\mathbf N-1}(x)
-A_{2\mathbf N-2}(x)+\displaystyle\frac{1}{2}\Sigma_{\mathbf N}^2-\Pi_{\mathbf N}=0.
\end{equation}

\item[\text{\rm{(b)}}]  Next, we express the constraint in \eqref{4.14} in terms of 
$\det[\mathbf M(x)]$ and $\det[\mathbf M_2(x)].$ This is done by using \eqref{4.7}
and \eqref{4.8} in \eqref{4.14}. We get
\begin{equation}\label{4.15}
\displaystyle\frac{1}{2} \left(\Sigma_{\mathbf N}
+\displaystyle\frac{(\det[\mathbf M(x)])'}{\det[\mathbf M(x)]}\right)^2
+\displaystyle\frac{1}{2}\left(\displaystyle\frac{(\det[\mathbf M(x)])'}{\det[\mathbf M(x)]}\right)'
+\displaystyle\frac{\det[\mathbf M_2(x)]}{\det[\mathbf M(x)]}+\displaystyle\frac{1}{2}\Sigma_{\mathbf N}^2-\Pi_{\mathbf N}=0.
\end{equation}
Simplifying \eqref{4.15}, we obtain the equivalent constraint given by
\begin{equation}\label{4.16}
\left(\Sigma_{\mathbf N}^2-\Pi_{\mathbf N}
\right)\det[\mathbf M(x)]+\Sigma_{\mathbf N}(\det[\mathbf M(x)])'+\displaystyle\frac{1}{2}(\det[\mathbf M(x)])''+\det[\mathbf M_2(x)]=0.
\end{equation}

As indicated in Section~5 of \cite{AETU2026}, each of the 
quantities $\det[\mathbf M(x)],$ $\left(\det[\mathbf M(x)]\right)',$ $\left(\det[\mathbf M(x)]\right)'',$ and
$\det[\mathbf M_2(x)]$ is a polynomial of degree $2\mathbf N$ in the variables 
$\chi_1,$ $\chi_2,$ $\dots,$ $\chi_{\mathbf N},$ where each variable appears
in the polynomial at most to the second power. We recall that $\chi_j$ is the quantity 
defined in \eqref{2.2}, and hence the $x$-derivatives $\chi_j'$ and $\chi_j''$
are given by 
 \begin{equation}\label{4.17}
\chi_j'=-(k_j+k_j^\ast)\,\chi_j, \quad \chi_j''=(k_j+k_j^\ast)^2\,\chi_j, \qquad 1\le j\le\mathbf N.
\end{equation}

 \item[\text{\rm{(c)}}] Since the complex-valued constants $k_j$ are distinct
and they appear in the input data set symmetrically, the constraint in \eqref{4.16} allows us 
to determine the corresponding constraint on each of the
modified dependency constants $\gamma_j$ for $1\le j\le\mathbf N.$ For this, we
proceed as follows. By
letting $\gamma_l=0$ for $l\neq j,$ as in Section~5 of \cite{AETU2026}, we obtain
 \begin{equation}\label{4.18}
\det[\mathbf M(x)]=\nu+\left[\nu_{j0}\gamma_j+\nu_{0j}\gamma_j^\ast\right]\chi_j+\nu_{jj}|\gamma_j |^2\,\chi_j^2,
\end{equation}  
where we have defined
\begin{equation}\label{4.19}
\nu:=\det\begin{bmatrix}
k_1^{2\mathbf N-1} & k_1^{2\mathbf N-2}& \cdots &k_1 &1\\
\noalign{\medskip}
(k_1^\ast)^{2\mathbf N-1}&(k_1^\ast)^{2\mathbf N-2} & \cdots & k_1^\ast & 1\\
\noalign{\medskip}
\vdots & \vdots & \ddots & \vdots & \vdots \\
\noalign{\medskip}
k_{\mathbf N} ^{2\mathbf N-1}& k_{\mathbf N} ^{2\mathbf N-2}& \cdots & k_{\mathbf N}& 1\\
\noalign{\medskip}
(k_{\mathbf N}^\ast) ^{2\mathbf N-1}& (k_{\mathbf N}^\ast) ^{2\mathbf N-2}& \cdots & k_{\mathbf N}^\ast & 1
\end{bmatrix},
\end{equation}
\begin{equation*}
\nu_{j0}:=\nu\bigg|_{k_j\mapsto zk_j},\quad
\nu_{0j}:=\nu\bigg|_{k^\ast_j\mapsto z^2k^\ast_j},\quad
\nu_{jj}:=\nu\bigg|_{\substack{k_j\mapsto zk_j\\
k^\ast_j\mapsto z^2k^\ast_j}}.
\end{equation*}
It is known \cite{AETU2026} that $\nu$ in \eqref{4.19} can be expressed in the form of a product as
\begin{equation*}
\nu=[(k_1-k_1^\ast)(k_1-k_2)\cdots(k_1-k^\ast_{\mathbf N})][(k^\ast_1-k_2)\cdots(k^\ast_1-k^\ast_{\mathbf N})]
\cdots[(k_{\mathbf N}-k^\ast_{\mathbf N})].
\end{equation*}
Using \eqref{4.17} in \eqref{4.18}, we have
\begin{equation}\label{4.22}
(\det[\mathbf M(x)])'=
-(k_j+k_j^\ast)\, [\nu_{j0}\,\gamma_j+\nu_{0j}\,\gamma_j^\ast]\,\chi_j-2(k_j+k_j^\ast)\,\nu_{jj}\,\gamma_j\,\gamma_j^\ast\,\chi_j^2,
\end{equation}
\begin{equation}\label{4.23}
(\det[\mathbf M(x)])''=
(k_j+k_j^\ast)^2\, [\nu_{j0}\,\gamma_j+\nu_{0j}\,\gamma_j^\ast]\,\chi_j+4(k_j+k_j^\ast)^2\,\nu_{jj}\,\gamma_j\,\gamma_j^\ast\,\chi_j^2.
\end{equation}
In a way similar to \eqref{4.18}, by letting $\gamma_l=0$ for $l\neq j,$ with the help of \eqref{4.1} and \eqref{4.3}, we get
 \begin{equation}\label{4.24}
\det[\mathbf M_2(x)]=\mu+\left[\mu_{j0}\gamma_j+\mu_{0j}\gamma_j^\ast\right]\chi_j+\mu_{jj}|\gamma_j |^2\,\chi_j^2.
\end{equation} 
The constants $\mu,$ $\mu_{j0},$ $\mu_{0j},$ $\mu_{jj}$ can be evaluated as described in Section~5 of \cite{AETU2026}, and we have
\begin{equation}\label{4.25}
\mu=(\Pi_{\mathbf N}-\Sigma^2_{\mathbf N})\,\nu,
\end{equation}
\begin{equation}\label{4.26}
\mu_{j0}:=\mu\bigg|_{k_j\mapsto zk_j},\quad
\mu_{0j}:=\mu\bigg|_{k^\ast_j\mapsto z^2k^\ast_j},\quad
\mu_{jj}:=\mu\bigg|_{\substack{k_j\mapsto zk_j\\
k^\ast_j\mapsto z^2k^\ast_j}},
\qquad 1\le j\le\mathbf N.
\end{equation}  
  
   \item[\text{\rm{(d)}}] We remark that the constants $\nu,$ $\nu_{j0},$ $\nu_{0j},$ $\mu,$ $\mu_{j0},$ $\mu_{0j}$ 
possess various symmetries that can be verified
by using some elementary properties of determinants. For example, we have
\begin{equation}\label{4.27}
\begin{cases}
\nu^\ast=(-1)^{\mathbf N}\nu,\\
\nu_{j0}^\ast=(-1)^{\mathbf N}\nu_{0j}, \qquad 1\le j\le \mathbf N,\\
\nu_{0j}^\ast=(-1)^{\mathbf N}\nu_{j0},\qquad 1\le j\le\mathbf N.
\end{cases}
\end{equation}
The constants $\Sigma_{\mathbf N}$ and $\Pi_{\mathbf N}$ defined in \eqref{4.9} and 
\eqref{4.13}, respectively, are real valued, and hence we have
\begin{equation}\label{4.28}
\Sigma_{\mathbf N}^\ast=\Sigma_{\mathbf N}, \quad  \Pi_{\mathbf N}^\ast=\Pi_{\mathbf N}.
\end{equation}
We remark that \eqref{4.28} follows from the fact that each pair $k_j$ and $k_j^\ast$ appear symmetrically in both 
$\Sigma_{\mathbf N}$ and $\Pi_{\mathbf N}.$
Thus, with the help of \eqref{4.27} and \eqref{4.28}, from \eqref{4.25} and \eqref{4.26} we conclude that
 \begin{equation}\label{4.29}
\begin{cases}
\mu^\ast=(-1)^{\mathbf N}\mu,\\
\mu_{j0}^\ast=(-1)^{\mathbf N}\mu_{0j}, \qquad 1\le j\le \mathbf N,\\
\mu_{0j}^\ast=(-1)^{\mathbf N}\mu_{j0},\qquad 1\le j\le\mathbf N.
\end{cases}
\end{equation}

\item[\text{\rm{(e)}}] Using \eqref{4.18} and \eqref{4.22}--\eqref{4.24} in the constraint given in 
\eqref{4.16}, we see that the constant term in the polynomial
appearing on the right-hand side of \eqref{4.16} must be equal to $0,$ which is equivalent to 
\eqref{4.25}. Similarly, the coefficient of $\chi_j$ on the left-hand side
of \eqref{4.16} must vanish, and hence we get
\begin{equation}\label{4.30}
\epsilon_j\left(\nu_{j0}\,\gamma_j+\nu_{0j}\,\gamma_j^\ast\right)+\left(\mu_{j0}\,\gamma_j
+\mu_{0j}\,\gamma_j^\ast\right)=0, \qquad 1\le j \le \mathbf N,
\end{equation}
where we have defined
\begin{equation}\label{4.31}
\epsilon_j:=\Sigma_{\mathbf N}^2-\Pi_{\mathbf N}-(k_j+k_j^\ast)\,\Sigma_{\mathbf N}
+\displaystyle\frac{1}{2}\,(k_j+k_j^\ast)^2, \qquad 1\le j \le \mathbf N.
\end{equation}
Using \eqref{4.28} in \eqref{4.31}, we observe the constant $\epsilon_j$ is real valued. 
The equality in \eqref{4.30} yields the constraint on $\gamma_j$ for which
we have been looking.

\item[\text{\rm{(f)}}] The constraint on $\gamma_j$ expressed in \eqref{4.30} can be further 
evaluated by writing $\gamma_j$ in terms of the real and imaginary parts
of $\gamma_j,$ which we denote by $r_j$ and $s_j,$ respectively. Thus, we let
\begin{equation}\label{4.32}
\gamma_j=r_j+is_j,\quad  \gamma_j^\ast=r_j-is_j.
\end{equation}
Using \eqref{4.27} and \eqref{4.29} in \eqref{4.30}, we write \eqref{4.30} in the equivalent form as
\begin{equation*}
\epsilon_j\left[\nu_{j0}\gamma_j+(-1)^{\mathbf N}\nu_{j0}^\ast\gamma_j^\ast\right]+\left[\mu_{j0}\gamma_j+(-1)^{\mathbf N}\mu_{j0}\gamma_j^\ast\right]=0, \qquad 1\le j \le \mathbf N.
\end{equation*}
Since $\epsilon_j$ is a real-valued constant, we see that \eqref{4.31} is equivalent to
\begin{equation*}
\begin{cases}
\epsilon_j\, \text{\rm{Re}}[\nu_{j0}\gamma_j]+ \text{\rm{Re}}[\mu_{j0}\gamma_j]=0, \qquad 
\text{\rm{even }}
\mathbf N,\\
\epsilon_j\, \text{\rm{Im}}[\nu_{j0}\gamma_j]+ \text{\rm{Im}}[\mu_{j0}\gamma_j]=0, \qquad \text{\rm{odd }}\mathbf N.
\end{cases}
\end{equation*}
Using \eqref{4.32} in \eqref{4.30}, we express the constraint on $\gamma_j$ in terms of $r_j$ and $s_j$ as
\begin{equation}\label{4.35}
\epsilon_j\left[(r_j+is_j)\,\nu_{j0}+(r_j-is_j)\,\nu_{0j}\right]+\left[(r_j+is_j)\,\mu_{j0}+(r_j-is_j)\,\mu_{0j}\right]=0, \qquad 1\le j \le \mathbf N.
\end{equation}
We note that \eqref{4.35} is equivalent to
\begin{equation}\label{4.36}
s_j=ir_j\,\displaystyle\frac{\epsilon_j(\nu_{j0}+\nu_{0j})+(\mu_{j0}
+\mu_{0j})}{\epsilon_j(\nu_{j0}-\nu_{0j})+(\mu_{j0}-\mu_{0j})}, \qquad 1\le j \le \mathbf N.
\end{equation}
Using \eqref{4.36} in the first equality in \eqref{4.32}, we express the constraint on $\gamma_j$ as
\begin{equation*}
\gamma_j=r_j\left[1-\displaystyle\frac{\epsilon_j(\nu_{j0}+\nu_{0j})
+(\mu_{j0}+\mu_{0j})}{\epsilon_j(\nu_{j0}-\nu_{0j})+(\mu_{j0}-\mu_{0j})}\right], \qquad 1\le j \le \mathbf N,
\end{equation*}
which is equivalent to
\begin{equation*}
\gamma_j=-\displaystyle\frac{2r_j(\epsilon_j\nu_{0j}+\mu_{0j})}{\epsilon_j(\nu_{j0}-\nu_{0j})
+(\mu_{j0}-\mu_{0j})}, \qquad 1\le j \le \mathbf N,
\end{equation*}
with the real-valued constants $r_j$ being arbitrary.
Consequently, using the input data set $\{k_j,r_j\}_{j=1}^{\mathbf N}$
yields the $\mathbf N$-soliton solution $Q$ to the Kaup--Kupershmidt equation
\eqref{1.1}. However, there is a slight drawback of using $\{k_j,r_j\}_{j=1}^{\mathbf N}$
as the input data set. The  domains of the real $r_j$-values must be restricted appropriately
to avoid the $\mathbf N$-soliton solution $Q$ to have singularities at certain $(x,t)$-values.
Thus, we prefer to use the input data set $\{k_j,c_j\}_{j=1}^{\mathbf N},$
where the only restriction on each arbitrary parameter $c_j$ is the positivity so that
the corresponding $Q$ does not have any singularities when $x$ and $t$ take all real values.

 \item[\text{\rm{(g)}}] We write the constraint in \eqref{4.30} involving $\gamma_j$ and $\gamma_j^\ast$ as
 \begin{equation*}
(\epsilon_j \,\nu_{j0}+\mu_{j0})\,\gamma_j+(\epsilon_j \,\nu_{0j}+\mu_{0j})\,\gamma_j^\ast=0,\qquad 1\le j \le \mathbf N,
\end{equation*}
from which we obtain
 \begin{equation}\label{4.40}
\gamma_j^\ast=-\displaystyle\frac{\epsilon_j \,\nu_{j0}+\mu_{j0}}{\epsilon_j \,\nu_{0j}+\mu_{0j}}\,\gamma_j,\qquad 1\le j \le \mathbf N.
\end{equation}
Hence, the $\mathbf N$-soliton solution to the Kaup--Kupershmidt equation \eqref{1.1} 
is obtained from the $\mathbf N$-soliton solution to \eqref{3.7} by using
the restriction specified in \eqref{4.40}. From \eqref{4.4}, we see that this task can be 
accomplished by imposing the constraint \eqref{4.40} on $\det[\mathbf M(x)].$
We recall that our goal is to express the $\mathbf N$-soliton solution to \eqref{1.1} in terms 
of the input data set $\{k_j,c_j\}_{j=1}^{\mathbf N}.$ Hence, we need
to relate each $c_j$ to $\gamma_j$ satisfying the constraint \eqref{4.40}.

 \item[\text{\rm{(h)}}] We refer to the arbitrary positive constant $c_j$ appearing in the 
$\mathbf N$-soliton solution formula \eqref{2.16} as the bound-state
constant at $k=k_j.$ The relationship between $k_j$ and $\gamma_j$ with the restriction 
\eqref{4.40} is seen as follows. By comparing \eqref{2.16} with \eqref{4.4},
we see that $\tau$ in \eqref{2.16} and $\det[\mathbf M(x)]$ in \eqref{4.4} are related to each other as
\begin{equation}\label{4.41}
\nu\,\tau=\det[\mathbf M(x)],
\end{equation}
where we recall that $\nu$ is the constant in \eqref{4.19}. Letting $\chi_l=0$ for 
$l\neq j$ in \eqref{4.41}, with the help of \eqref{2.17},  \eqref{2.29},  \eqref{2.35}, and
\eqref{4.18} we obtain
\begin{equation}\label{4.42}
\nu\left[1+2c_j\,\chi_j+\displaystyle\frac{1}{4}c_j^2\,\chi_j^2\right]=\nu+\left[\nu_{j0}
\,\gamma_j+\nu_{0j\,}\gamma_j^\ast\right]\chi_j+\nu_{jj}\,|\gamma_j|^2\chi_j^2.
\end{equation}
Since the $k_j$-values are distinct and \eqref{4.42} must be satisfied for all 
$x\in\mathbb R$ and $t\in\mathbb R,$ we see that the constant terms, the coefficients of $\chi_j,$ and
the coefficients of $\chi_j^2$ on both sides of \eqref{4.42} must
respectively agree.
By comparing the coefficients of $\chi_j$ and
$\chi_j^2$ on each side of \eqref{4.42}, we see that we must have
\begin{equation}\label{4.43}
2\,\nu c_j=\nu_{j0}\,\gamma_j+\nu_{0j}\,\gamma_j^\ast, \qquad 1\le j \le \mathbf N,
\end{equation}
\begin{equation}\label{4.44}
\displaystyle\frac{1}{4}\,\nu\, c_j^2=\nu_{jj} \,|\gamma_j|^2, \qquad  1\le j \le \mathbf N.
\end{equation}
From \eqref{4.43}, we observe that the value of each arbitrary positive $c_j$ in 
the data set $\{k_j,c_j\}_{j=1}^{\mathbf N}$ in the $\mathbf N$-soliton solution formula
\eqref{2.16} is given by
\begin{equation*}
c_j=\displaystyle\frac{\nu_{j0}\,\gamma_j+\nu_{0j}\,\gamma_j^\ast}{2\nu},\qquad 1\le j \le \mathbf N,
\end{equation*}
where $\gamma_j$ and $\gamma_j^\ast$ are related to each other by the constraint \eqref{4.40}. In other words, we have
\begin{equation*}
c_j=\displaystyle\frac{1}{2\nu}\left[\nu_{j0}-\nu_{0j}\displaystyle\frac{\epsilon_j\,\nu_{j0}+\mu_{j0}}{\epsilon_j\,\nu_{0j}
+\mu_{0j}}\right]\gamma_j,\qquad 1\le j \le \mathbf N,
\end{equation*}
which is equivalent to the simplified expression
\begin{equation}\label{4.47}
c_j=\displaystyle\frac{\gamma_j(\nu_{j0}\,\mu_{0j}-\nu_{0j}\,\mu_{j0})}{2\nu(\epsilon_j\,\nu_{0j}+\mu_{0j})},
\qquad 1\le j \le \mathbf N.
\end{equation}
Using the fact that $\epsilon_j$ is real and we have the symmetry properties listed in \eqref{4.27} and 
\eqref{4.29}, we can write \eqref{4.47} in the equivalent form as
\begin{equation}\label{4.48}
\gamma_j=\displaystyle\frac{2\nu\,(\epsilon_j\,\nu_{j0}^\ast+\mu_{j0}^\ast)}{\nu_{j0}\,
\mu_{j0}^\ast-\nu_{j0}^\ast\,\mu_{j0}}\,c_j,\qquad 1\le j \le \mathbf N.
\end{equation}
By taking the complex conjugate, from \eqref{4.48} we get
\begin{equation}\label{4.49}
\gamma_j^\ast=\displaystyle\frac{2\nu^\ast\,(\epsilon_j\,\nu_{j0}+\mu_{j0})}{\nu_{j0}^\ast\,
\mu_{j0}-\nu_{j0}\,\mu_{j0}^\ast}\,c_j,\qquad 1\le j \le \mathbf N.
\end{equation}
Using \eqref{4.48} and \eqref{4.49}, we replace the input data set
$\{k_j,\gamma_j\}_{j=1}^{\mathbf N}$ with the data set 
the data set 
$\{k_j,c_j\}_{j=1}^{\mathbf N},$ 
and in turn with the data set 
$\{p_j,c_j\}_{j=1}^{\mathbf N},$
with the distinct $p_j$ values as in \eqref{2.6} and
with the $c_j$-values being arbitrary positive constants.
Using the restrictions \eqref{4.48} and \eqref{4.49} in \eqref{4.41}, we then get
the $\mathbf N$-soliton solution formula \eqref{2.16} for \eqref{1.1}.
It is straightforward but lengthy to show that \eqref{4.44} is satisfied
when \eqref{4.48} and \eqref{4.49} hold. The proof can be obtained by using mathematical induction on $\mathbf N.$

\end{enumerate}

\section{The Marchenko integral equation}
\label{section5}

In this section we present the construction of the $\mathbf N$-soliton solution formula \eqref{2.16} with the input data set
$\{p_j,c_j\}_{j=1}^{\mathbf N},$ where we recover the time-evolved potential $Q$ from the solution to the relevant Marchenko integral equation. In 
Section~6 of \cite{AETU2026} the reader can find the derivation of and the solution to the Marchenko integral equation associated with the inverse
scattering problem for \eqref{3.1} in the reflectionless case using the input data set $\{k_j,\gamma_j\}_{j=1}^{\mathbf N}.$ Furthermore, the
$\mathbf N$-soliton solution formula for the integrable system \eqref{3.7} with the input data set $\{k_j,\gamma_j\}_{j=1}^{\mathbf N}$ is already
available in Section~6 of \cite{AETU2026}, where the time-evolved potentials $Q$ and $P$ are recovered from the solution to the
aforementioned Marchenko integral equation. Hence, instead of showing the details of the derivation of the Marchenko integral equation and its
solution with the input data set $\{k_j,\gamma_j\}_{j=1}^{\mathbf N},$ we can directly use the formula for $Q$ obtained from the solution to that
Marchenko integral equation. Then, by using the restrictions \eqref{4.48} and \eqref{4.49}, we can obtain the $\mathbf N$-soliton solution formula
\eqref{2.16} by expressing $Q$ in terms of the input data set $\{p_j,c_j\}_{j=1}^{\mathbf N}$ with the distinct $p_j$-values as in \eqref{2.6} and with
the $c_j$-values being arbitrary positive parameters.

We first describe the recovery of the time-evolved $Q$ appearing in the integrable system \eqref{3.7} 
by using the input data set $\{k_j,\gamma_j\}_{j=1}^{\mathbf N}.$ We refer the
reader to Section~6 of \cite{AETU2026} for the details. Our procedure is outlined in the following steps:

\begin{enumerate}
    
\item[\text{\rm{(i)}}] The Marchenko integral equation corresponding to the input data set $\{k_j,\gamma_j\}_{j=1}^{\mathbf N}$ is given by (6.27)
of \cite{AETU2026}, i.e. we have
\begin{equation}\label{5.1}
\hat G(x,y)+\Omega(0,y)+\int_{-\infty}^0 d\zeta\,\hat G(x,\zeta)\,\Omega(\zeta,y)=0,\qquad y<0,
\end{equation}
where the nonhomogeneous term $\Omega(0,y)$ is obtained from the integral kernel $\Omega(\zeta,y)$ by letting $\zeta=0.$
The integral kernel $\Omega(\zeta,y)$ in \eqref{5.1} is expressed in terms of the input data set, and we have
\begin{equation}\label{5.2}
\Omega(\zeta,y)=\displaystyle\sum_{j=1}^{\mathbf N} \left(b_j
\,e^{ik_j\zeta-iz^2k_j y}-z\, b_j^\ast\,e^{i z k_j^\ast \zeta-iz^2k_j^\ast y}\right),
\end{equation}
where $b_j$ and its complex conjugate $b_j^\ast$ are given by
\begin{equation}
\label{5.3}
b_j:=\displaystyle\frac{iz^2\,\Gamma(-z k_j)\,\gamma_j\,\chi_j}{(k_j-k_j^\ast)\,\Gamma_j(-k_j)},\quad b_j^\ast
=\displaystyle\frac{iz\,\Gamma(-z^2 k_j^\ast)\,\gamma_j^\ast\,\chi_j}{(k_j-k_j^\ast)\,\Gamma_j(-k_j^\ast)},\qquad
1\le j\le\mathbf N.
\end{equation}
We recall that $z$ is the special constant in \eqref{2.4}, the function $\Gamma(k)$ is defined in \eqref{3.13}, 
and the quantity $\chi_j$ is as in \eqref{2.7}.  The quantity $\Gamma_j(k)$ is related to 
the function $\Gamma(k),$ and it is defined as
\begin{equation*}
\Gamma_j(k):=\displaystyle\frac{\Gamma(k)}{(k+k_j)(k+k_j^\ast)},\qquad 1\le j\le\mathbf N.
\end{equation*}

 \item[\text{\rm{(ii)}}] Using the input data set $\{k_j,\gamma_j\}_{j=1}^{\mathbf N},$ we construct the 
Marchenko kernel $\Omega(\zeta,y)$ and the
nonhomogeneous term $\Omega(0,y)$. As seen from \eqref{5.2}, the resulting kernel $\Omega(\zeta,y)$ 
is separable in $\zeta$ and $y,$
and hence the Marchenko integral equation \eqref{5.1} can be solved explicitly by using methods from 
linear algebra. For the derivation of the
corresponding solution $\hat G(x,y),$ we refer the reader to Section~6 of \cite{AETU2026}. We have
\begin{equation}\label{5.5}
\hat G(x,y)=\displaystyle\frac{\det\begin{bmatrix}\begin{array}{c | c c c c c} 
	0 & e^{-iz^2k_1 y}&e^{-iz^2k_1^\ast y}&\cdots&
e^{-iz^2k_{\mathbf N} y}&e^{-iz^2k_{\mathbf N}^\ast y}\\ 
	\hline 
	\mathbf b(x)& && \mathbf m(x)& &
\end{array}
\end{bmatrix}}{\det[\mathbf m(x)]},
\end{equation}
where the $2\mathbf N\times 2\mathbf N$ matrix $\mathbf m(x)$ and the column vector 
$\mathbf b(x)$ with $2\mathbf N$ components are given by
\begin{equation}\label{5.6}
\begin{split}\mathbf m(x)&
:=\\
&\begin{bmatrix}
1+\frac{i z b_1}{k_1^\ast+k_1} & \frac{-i b_1}{k_1+k_1} & \frac{i z b_1}{k_1^\ast+k_2} & 
\frac{-i b_1}{k_1+k_2} & \cdots & \frac{i z b_1}{k_1^\ast+k_{\mathbf N}} & \frac{-i b_1}{k_1+k_{\mathbf N}}\\
\noalign{\medskip}
\frac{-i z^2 b_1^\ast}{k_1+ k_1} & 1+\frac{-iz^2 b_1^\ast}{k_1+k_1^\ast}&\frac{-iz^2 
b_1^\ast}{k_1+ k_2} & \frac{-iz^2  b_1^\ast}{k_1+ k_2^\ast} & \cdots & \frac{-i z^2 
b_1^\ast}{k_1+k_{\mathbf N}} &\frac{-iz^2  b_1^\ast}{k_1+ k_{\mathbf N}^\ast}\\
\noalign{\medskip}
\frac{i z b_2}{k_2^\ast+k_1} &\frac{-i b_2}{k_2+k_1} & 1+\frac{i z b_2}{k_2^\ast+k_2} & 
\frac{-i b_2}{k_2+k_2} & \cdots & \frac{i z b_2}{k_2^\ast+k_{\mathbf N}} & \frac{-i b_2}{k_2+k_{\mathbf N}}\\
\noalign{\medskip}
\frac{-i z^2 b_2^\ast}{k_2+k_1} &\frac{-i z^2 b_2^\ast}{k_2+k_1^\ast} &\frac{-i z^2 b_2^\ast}{k_2+k_2} 
&1+\frac{-i z^2 b_2^\ast}{k_2+ k_2^\ast} & \cdots & \frac{-i z^2 b_2^\ast}
{k_2+k_{\mathbf N}} &\frac{-i z^2 b_2^\ast}{k_2+k_{\mathbf N}^\ast}\\
\noalign{\medskip}
\vdots&\vdots&\vdots&\vdots&\ddots&\vdots&\vdots\\
\noalign{\medskip}
\frac{i z b_{\mathbf N}}{k_{\mathbf N}^\ast+k_1} &\frac{-i b_{\mathbf N}}{k_{\mathbf N}
+k_1} & \frac{i z b_{\mathbf N}}{k_{\mathbf N}^\ast+k_2}&\frac{-i b_{\mathbf N}}
{k_{\mathbf N}+k_2} & \cdots & 1+\frac{i z b_{\mathbf N}}{k_{\mathbf N}^\ast+
k_{\mathbf N}} & \frac{-i b_{\mathbf N}}{k_{\mathbf N}
+k_{\mathbf N}}\\
\noalign{\medskip}
\frac{-i z^2  b_{\mathbf N}^\ast}{k_{\mathbf N}+ k_1}&\frac{-i z^2 b_{\mathbf N}^\ast}
{k_{\mathbf N}+k_1^\ast} & \frac{-i z^2 b_{\mathbf N}^\ast}{k_{\mathbf N}+k_2}& 
\frac{-i z^2 b_{\mathbf N}^\ast}{k_{\mathbf N}+k_2^\ast} & \cdots & \frac{-i z^2  
b_{\mathbf N}^\ast}{k_{\mathbf N}+ k_{\mathbf N}} &1+ \frac{-i z^2 b_{\mathbf N}^\ast}{k_{\mathbf N}+k_{\mathbf N}^\ast}
\end{bmatrix},
\end{split}
\end{equation}
\begin{equation}\label{5.7}
\mathbf b(x):=\begin{bmatrix}
b_1\\
-z b_1^\ast\\
b_2\\
-z b_2^\ast\\
\vdots\\
b_{\mathbf N}\\
-z b_{\mathbf N}^\ast\\
\end{bmatrix}.
\end{equation}
From \eqref{5.5}--\eqref{5.7}, we observe that $\hat G(x,y)$ is explicitly expressed in terms of the parameters in the input data set
$\{k_j,\gamma_j\}_{j=1}^{\mathbf N}.$

\item[\text{\rm{(iii)}}] As seen from (6.29) of
\cite{AETU2026}, 
the time-evolved potential $Q$ appearing in \eqref{3.7} is obtained from the solution $\hat G(x,y)$ to the Marchenko integral
equation \eqref{5.1} as
\begin{equation*}
Q(x)=3iz\,\displaystyle\frac{d \hat G(x,0^-)}{dx},\qquad x\in\mathbb R.
\end{equation*}
As shown in Section~6 of \cite{AETU2026}, we have
\begin{equation}\label{5.9}
Q(x)=3\,\left(\displaystyle\frac{(\det[\mathbf m(x)])'}{\det[\mathbf m(x)]}\right)'.
\end{equation}
Since the matrix $\mathbf m(x)$ is explicitly determined by the input data set $\{k_j,\gamma_j\}_{j=1}^{\mathbf N},$ its determinant
$\det[\mathbf m(x)]$ is also explicitly determined by that input data set. By comparing \eqref{2.16} and \eqref{5.9}, with the help of 
\eqref{2.17} and \eqref{5.6} we see that we have
\begin{equation}\label{5.10}
\tau=\det[\mathbf m(x)].
\end{equation}
From \eqref{4.41} and \eqref{5.10}, we see that $\det[\mathbf M(x)]$ is related to 
$\det[\mathbf m(x)]$ as in (6.46) of \cite{AETU2026}, i.e. we have
\begin{equation}\label{5.11}
\det[\mathbf M(x)]=\nu\det[\mathbf m(x)].
\end{equation}

\end{enumerate}

So far, we have determined the $\mathbf N$-soliton solution for \eqref{3.7} by expressing
the time-evolved potential $Q$ there in terms of the input data set $\{k_j,\gamma_j\}_{j=1}^{\mathbf N}.$ Our goal is
to impose the constraint $P=Q_x/2$ in order to determine the restriction on the modified
dependency constants $\gamma_j$ so that the resulting $Q$ in \eqref{5.9} yields 
the $\mathbf N$-soliton solution formula \eqref{2.16} for 
the Kaup--Kupershmidt
equation \eqref{1.1}. In fact, from Section~\ref{section5} we already know that the constraint
$P=Q_x/2$ results in the restrictions \eqref{4.48} and \eqref{4.49}, where $\gamma_j$ and $\gamma_j^\ast$
are expressed in terms of the arbitrary positive parameter $c_j$ with the coefficients determined by the
set $\{k_l\}_{l=1}^{\mathbf N}.$ Thus, by using \eqref{4.48} and \eqref{4.49} in \eqref{5.3}, 
we first express $b_j$ and $b_j^\ast$ in terms of $c_j.$
We then use the resulting expressions in \eqref{5.6}. Consequently, we obtain
$\det[\mathbf m(x)]$ explicitly expressed in terms of the parameters in the data set
$\{k_j,c_j\}_{j=1}^{\mathbf N}.$ Then, using \eqref{3.11}, we express $\det[\mathbf m(x)]$ explicitly
in terms of the input data set $\{p_j,c_j\}_{j=1}^{\mathbf N}.$
Hence, from \eqref{5.10} we conclude that 
$\tau$ is explicitly determined by the input data set
 $\{p_j,c_j\}_{j=1}^{\mathbf N},$ as outlined in Section~\ref{section2} starting with \eqref{2.1}.

\section{The conclusion}
\label{section6}

In this paper we have presented the $\mathbf N$-soliton solution formula, for any positive integer $\mathbf N,$ for the Kaup--Kupershmidt
equation \eqref{1.1}. This is done with the help of the solution to the inverse scattering problem for the third-order equation \eqref{3.1} in the
reflectionless case by using the input data set $\{p_j,c_j\}_{j=1}^{\mathbf N}.$ We have shown that the distinct positive $p_j$-values are related
to the bound-state poles of a corresponding left transmission coefficient for \eqref{3.1} and the arbitrary positive $c_j$-values are related to the
bound-state dependency constants. We have indicated that the aforementioned inverse scattering problem can be solved either by using the
solution of a related Riemann--Hilbert problem or by using the solution of a related Marchenko integral equation.

The function $\tau$ appearing in the $\mathbf N$-soliton solution formula \eqref{2.16} contains
$2^{2\mathbf N}$ terms, and hence it is not practical or useful to display all those
terms when $\mathbf N$ is large. We have prepared Mathematica notebooks containing the
explicit expressions for the function $\tau$ corresponding to
the $\mathbf N$-soliton solutions for $1\le \mathbf N\le 10.$ In each Mathematica
notebook, the user can assign particular values to the parameters $p_j$ and $c_j$ for
$1\le j\le\mathbf N,$ display the corresponding particular $\mathbf N$-soliton solution $Q$ explicitly
as a function of $x$ and $t,$ plot snapshots of that particular solution at any specific $t$-value,
and animate the aforementioned particular solution $Q$ to observe the time evolution of 
the $\mathbf N$-soliton wave. 

\begin{figure}[!ht]
     \centering
\includegraphics[width=1.in,height=1in]{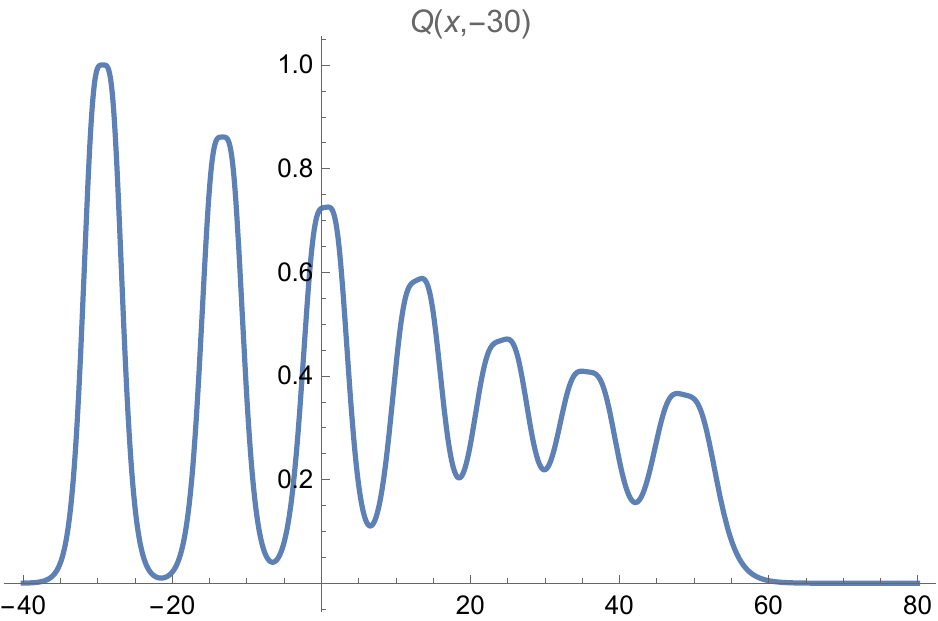}  \hskip .01in
\includegraphics[width=1.in,height=1in]{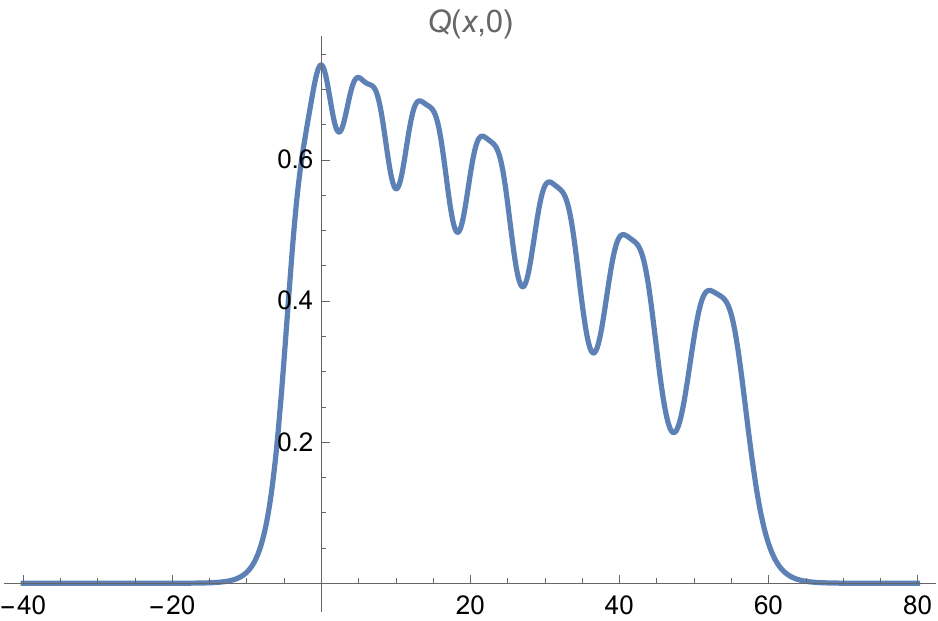} \hskip .01in
\includegraphics[width=1.in,height=1in]{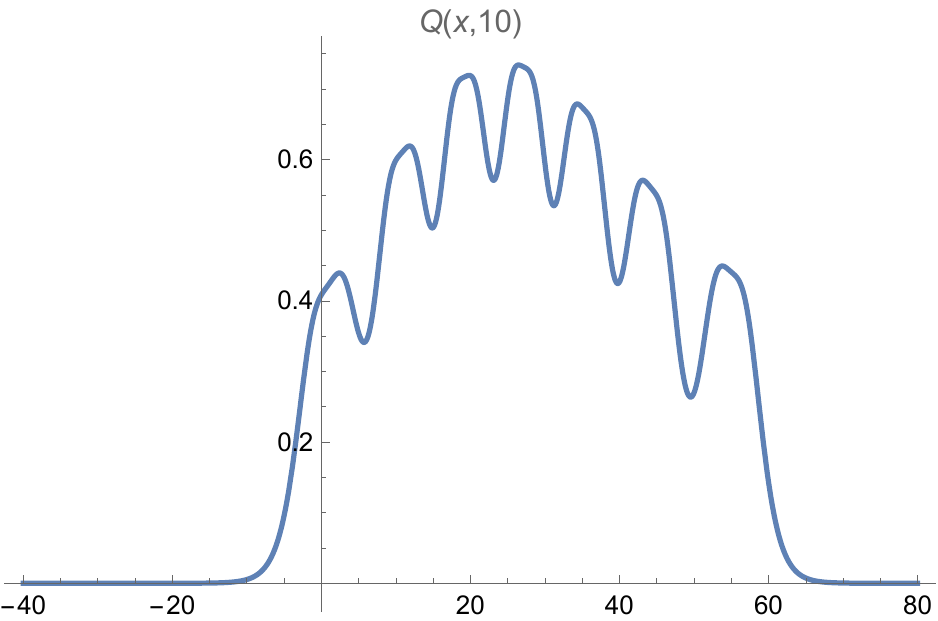}  \hskip .01in
\includegraphics[width=1.in,height=1in]{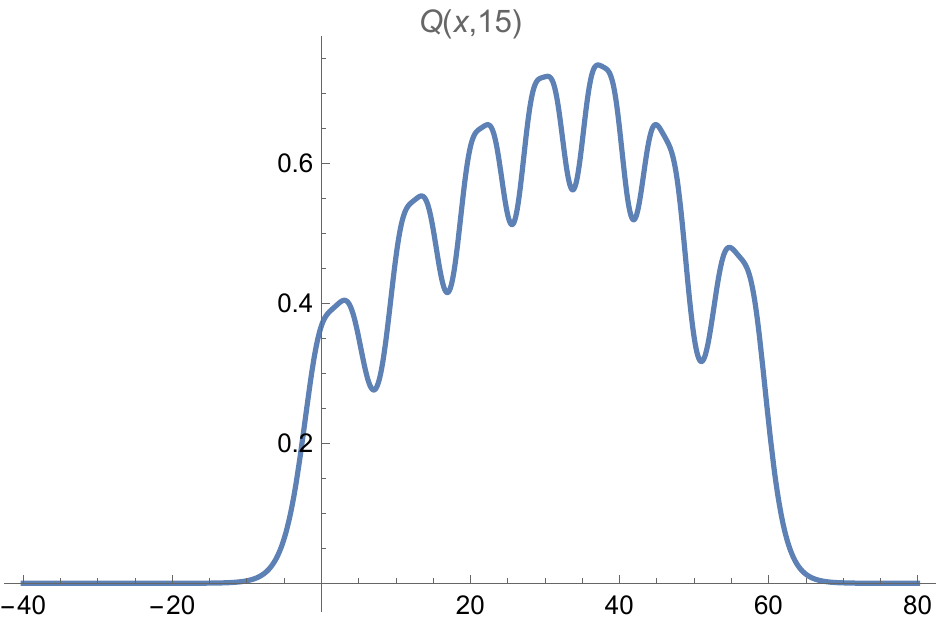}  \hskip .01in
\includegraphics[width=1.in,height=1in]{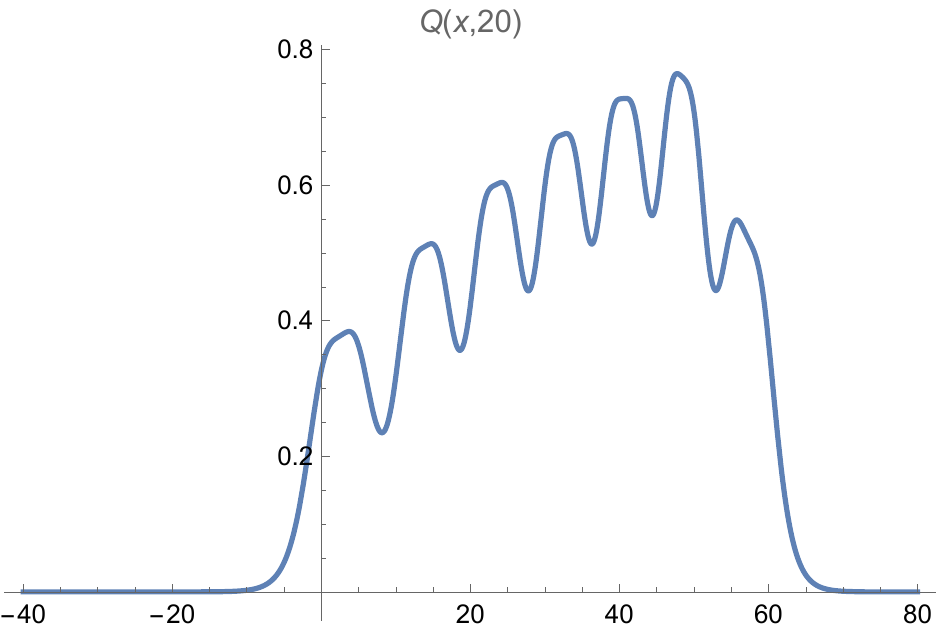}  \hskip .01in
\includegraphics[width=1.in,height=1in]{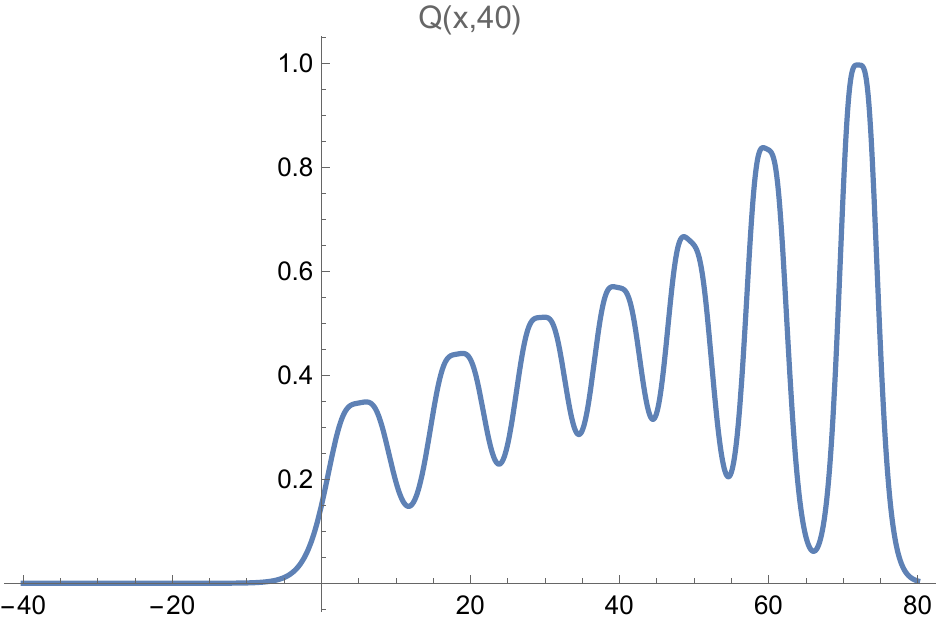}  
\caption{The snapshots for the $7$-soliton solution to the Kaup--Kupershmidt equation
with the input consisting of the $14$ parameters specified in
\eqref{6.1}.}
\label{figure6.1}
\end{figure}

In Figure~\ref{figure6.1}, as an example we provide the six snapshots for 
the $7$-soliton solution $Q(x,t)$ from
$t=-30$ to $t=40$ using the input data set $\{p_j,c_j\}_{j=1}^7$ with the $14$ parameters chosen as
\begin{equation}\label{6.1}
p_j=\displaystyle\frac{7+j}{14},\quad c_j=1,\qquad 1\le j\le 7.
\end{equation}

\begin{figure}[!ht]
     \centering
\includegraphics[width=1.in,height=1in]{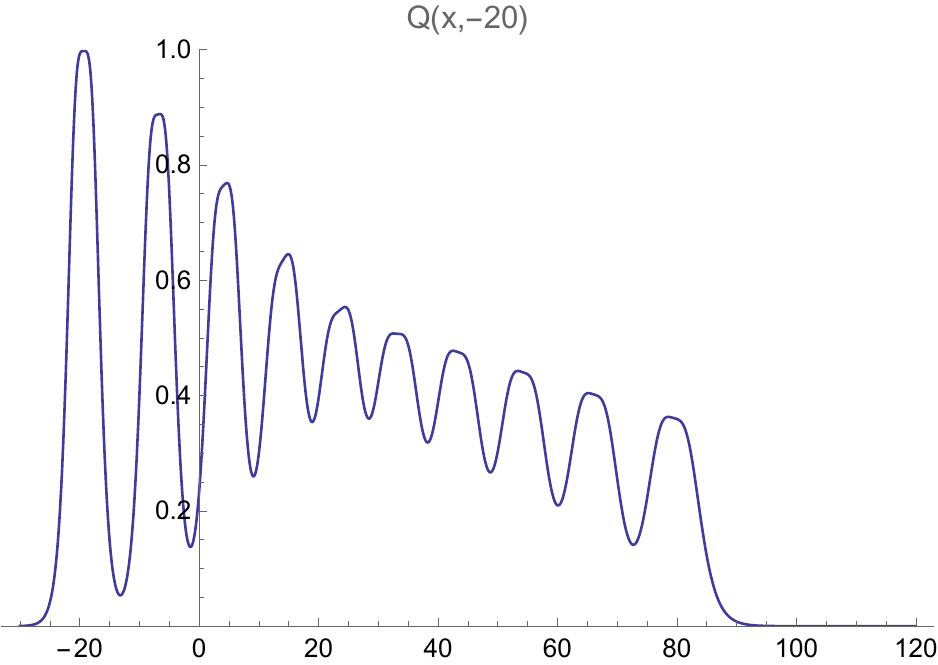}  \hskip .01in
\includegraphics[width=1.in,height=1in]{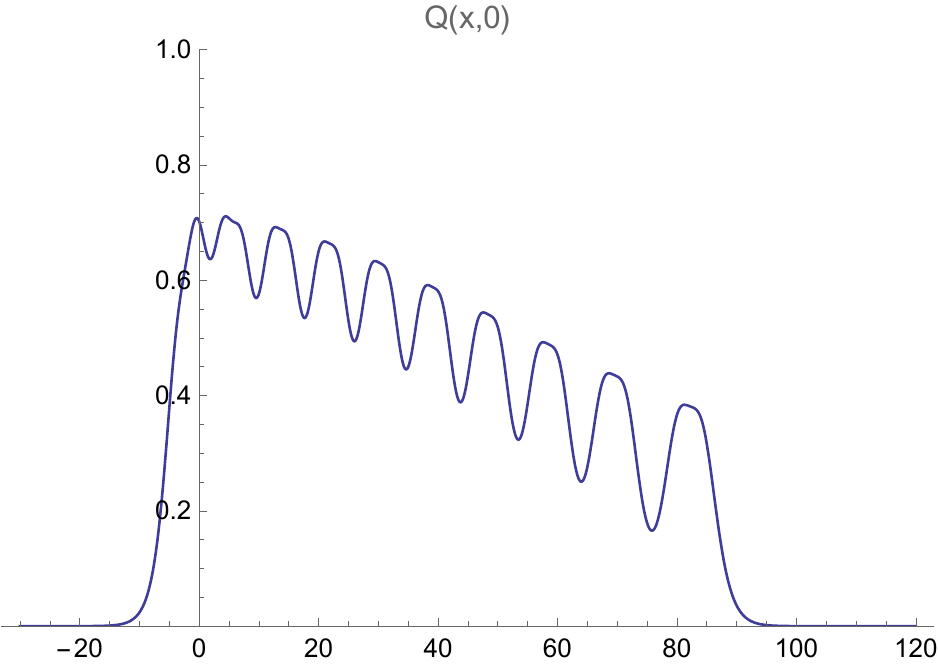} \hskip .01in
\includegraphics[width=1.in,height=1in]{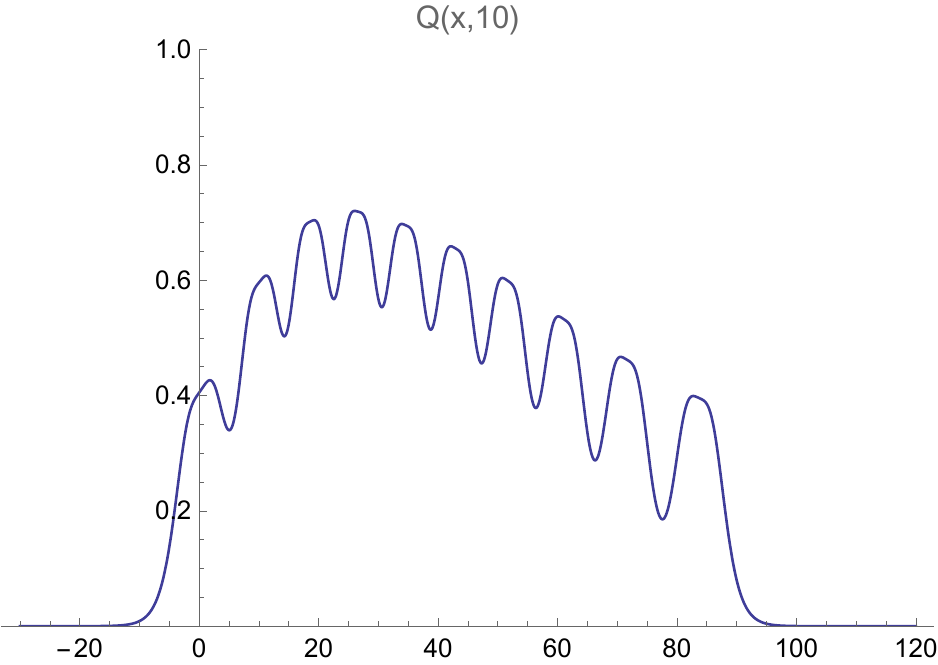}  \hskip .01in
\includegraphics[width=1.in,height=1in]{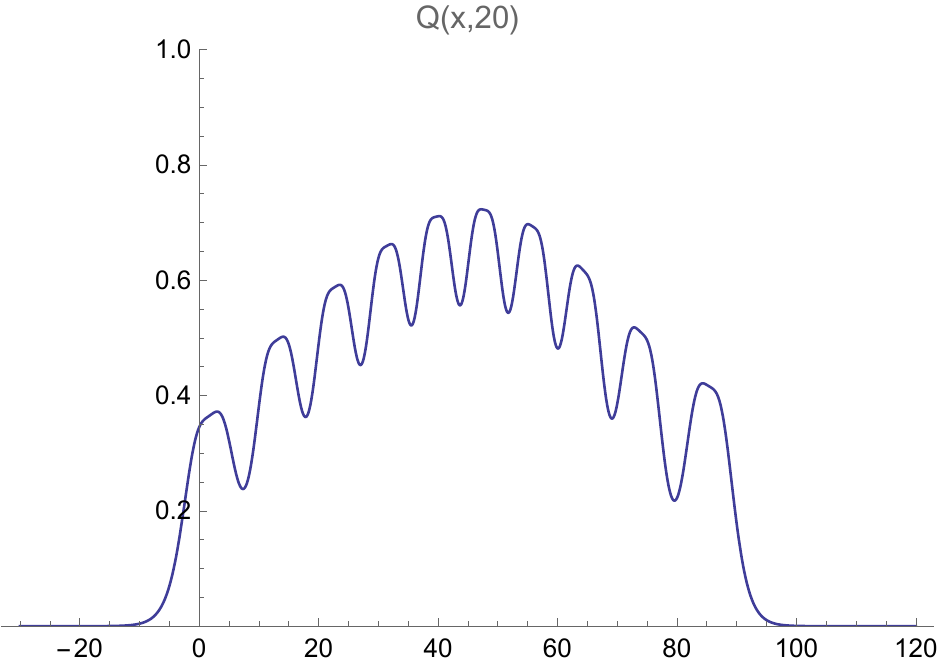}  \hskip .01in
\includegraphics[width=1.in,height=1in]{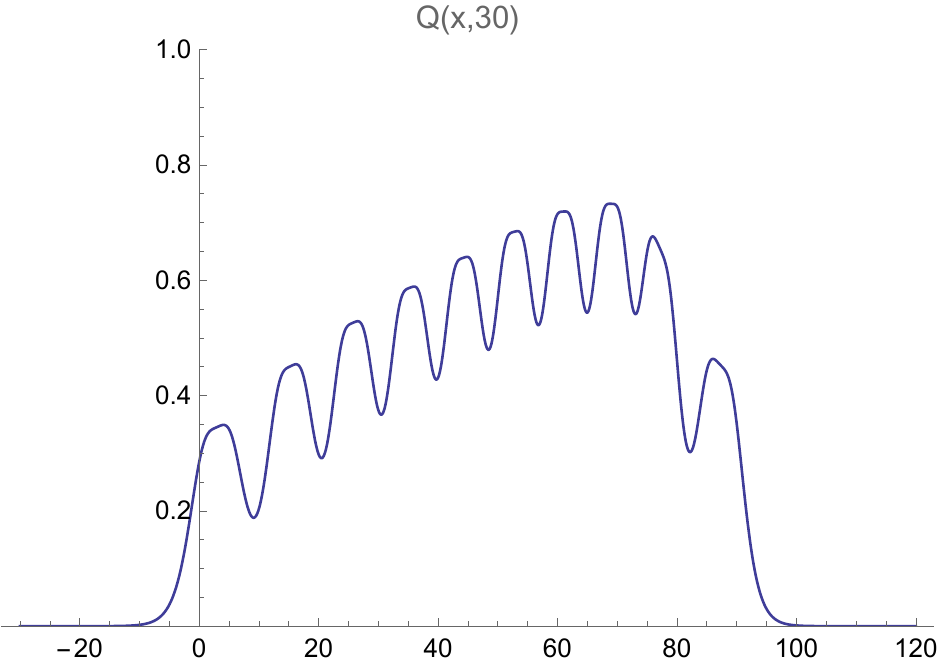}  \hskip .01in
\includegraphics[width=1.in,height=1in]{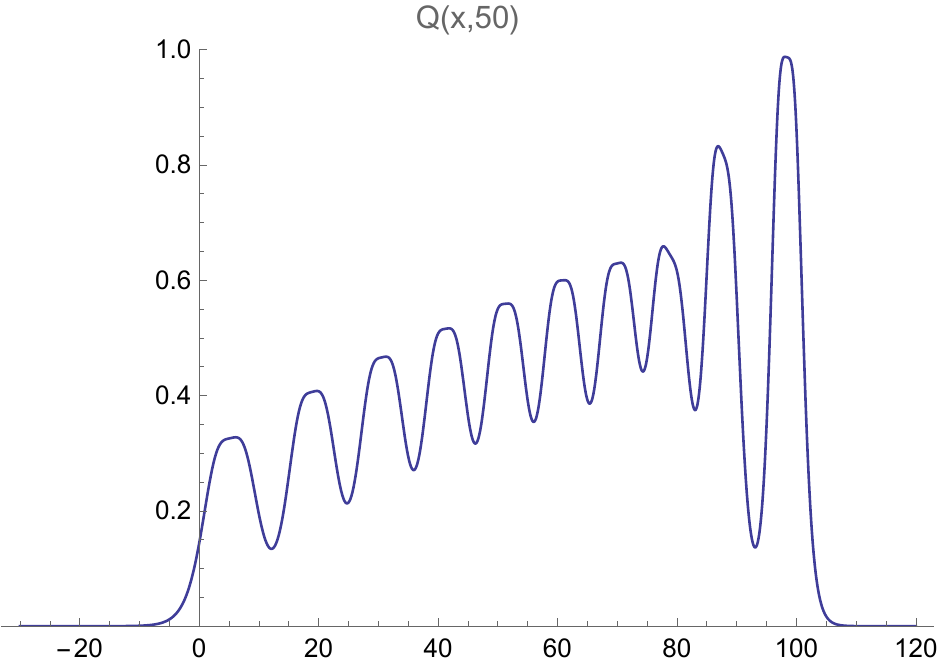}  
\caption{The snapshots for the $10$-soliton solution to the Kaup--Kupershmidt equation
with the input consisting of the $20$ parameters specified in
\eqref{6.2}.}
\label{figure6.2}
\end{figure}

 In Figure~\ref{figure6.2}, as another example we provide the six snapshots for the $10$-soliton solution $Q(x,t)$ from
$t=-20$ to $t=50$ using the input data set $\{p_j,c_j\}_{j=1}^{10}$ with the $20$ parameters chosen as
\begin{equation}\label{6.2}
p_j=\displaystyle\frac{10+j}{20},\quad c_j=1,\qquad 1\le j\le 10.
\end{equation}

The method presented here is not an ad-hoc method, but it is a fundamental procedure that also yields the $\mathbf N$-soliton solution formula,
for any positive integer $\mathbf N,$ for the Sawada--Kotera equation \eqref{1.2}. It also yields the complex-valued $\mathbf N$-soliton
solution formula, for any positive integer $\mathbf N,$ for the integrable system \eqref{3.7} by explicitly constructing the time-evolved potentials
$Q$ and $P$ in terms of the input data set $\{k_j,\gamma_j\}_{j=1}^{\mathbf N},$ where the parameters are related to the bound states for
\eqref{3.1} in the reflectionless case. For the aforementioned solution formulas and their derivations, we refer the reader to
\cite{ACTU2025,ACTU2026,AETU2026}.

The general technique we have used in this paper can readily be applied to other integrable nonlinear partial differential equations associated with
the third-order linear equation \eqref{3.1}. In the near future, using this technique
we plan to present the $\mathbf N$-soliton solution formulas, for any positive integer
$\mathbf N,$ for the bad Boussinesq equation \cite{B1872,H1973}
 \begin{equation*}
Q_{tt}-Q_{xx}+(Q^2)_{xx}-Q_{xxxx}=0,\qquad x,t\in\mathbb R,
\end{equation*}
the good Boussinesq equation \cite{BS1988,CLW2023}
 \begin{equation*}
Q_{tt}-Q_{xx}+(Q^2)_{xx}+Q_{xxxx}=0,\qquad x,t\in\mathbb R,
\end{equation*}
and the modified bad Boussinesq equation \cite{DTT1982,M1981}
 \begin{equation*}
Q_{tt}+(Q^2)_{xx}-Q_{xxxx}=0,\qquad x,t\in\mathbb R,
\end{equation*}
where the $\mathbf N$-soliton solution in each of these three cases is a real-valued function of $x$ and 
$t,$ vanishing as $x\to\pm\infty$ for each
fixed $t\in\mathbb R,$ and that the $2\mathbf N$ parameters in each soliton solution are related to the 
bound-state poles of a transmission
coefficient and the bound-state dependency constants associated with \eqref{3.1} in the 
reflectionless case with the appropriate choice of the two
potentials there.

\section*{Conflicts of Interest} The authors have no conflicts of interest to declare.

\end{document}